\documentclass[onecolumn,iop,apj]{emulateapj}   \newcommand{\rot}{}

\usepackage{graphicx}        
\usepackage{amssymb}         
\usepackage[usenames,dvipsnames]{color}          

\DeclareFontFamily{OT1}{pzc}{}
\DeclareFontShape{OT1}{pzc}{m}{it}%
             {<-> s * [1.1500] pzcmi7t}{}
\DeclareMathAlphabet{\mathscr}{OT1}{pzc}%
                                 {m}{it}

\usepackage{amssymb,amsmath}

\newcommand{\half}{{\textstyle\frac{1}{2}}}

\newcommand{\re}{\mathop{\rm Re}\nolimits}
\newcommand{\im}{\mathop{\rm Im}\nolimits}

\newcommand{\rd}{\mathrm{d}}
\newcommand{\rme}{e}

\newcommand{\deriv}[2]{\frac{\rd#1}{\rd#2}}

\newcommand{\e}{\hat{\mathbf{e}}}

\renewcommand{\k}{\mathbf{k}}

\newcommand{\boldd}{{\mathbf{d}}}
\newcommand{\boldu}{{\mathbf{u}}}
\newcommand{\boldv}{{\mathbf{v}}}
\newcommand{\boldw}{{\mathbf{w}}}

\newcommand{\A}{{\mathbf{A}}}
\newcommand{\B}{{\mathbf{B}}}
\newcommand{\E}{{\mathbf{E}}}
\newcommand{\F}{{\mathbf{F}}}
\newcommand{\I}{{\mathbf{I}}}
\newcommand{\X}{\mathbf{X}}

\renewcommand{\P}{{\mathbf{P}}}
\newcommand{\U}{\mathbf{U}}
\newcommand{\V}{\mathbf{V}}

\newcommand{\D}{\mbox{\boldmath$\mathsf{D}$}}

\newcommand{\Q}{\mbox{\boldmath$\mathsf{Q}$}}

\newcommand{\vdot}{{\boldsymbol{\cdot}}}
\newcommand{\vcross}{{\boldsymbol{\times}}}
\newcommand{\grad}{\mbox{\boldmath$\nabla$}}
\newcommand{\bxi}{\mbox{\boldmath$\xi$}}
\newcommand{\bkappa}{\mbox{\boldmath$\kappa$}}

\newcommand{\thth}{\hspace{1.5pt}}

\newcommand\Div{\grad\vdot\thth}

\newcommand{\kpar}{k_{\scriptscriptstyle\parallel}}

\newcommand{\ri}{{i}}

\newcommand{\calD}{{\mathcal{D}}}

\renewcommand{\leq}{\leqslant}  
  \renewcommand{\ge}{\geqslant}

\allowdisplaybreaks[1]

\begin{document}


\title{Benchmarking Fast-to-Alfv\'en Mode Conversion in a Cold MHD Plasma}

\author{Paul S.~Cally\altaffilmark{1} and
Shelley C.~Hansen}

\affil{Monash Centre for Astrophysics and School of
Mathematical Sciences,\\ Monash University, Clayton, Victoria 3800, Australia}
 \email{paul.cally@monash.edu}  \email{shelley.hansen@monash.edu} 
\altaffiltext{1}{The work presented here was largely carried out while PSC was visiting the High Altitude Observatory, National Center for Atmospheric Research, P.O. Box 3000, Boulder, CO 80307. The National Center for Atmospheric Research is sponsored by the National Science 
Foundation.}

\shortauthors{P.S. Cally and S.C. Hansen}

\shorttitle{Fast-to-Alfv\'en Mode Conversion}

\begin{abstract}
Alfv\'en waves may be generated via mode conversion from fast magneto-acoustic waves near their reflection level in the solar atmosphere, with implications both for coronal oscillations and for active region helio\-seismology. In active regions this reflection typically occurs high enough that the Alfv\'en speed $a$ greatly exceeds the sound speed $c$, well above the $a=c$ level where the fast and slow modes interact. In order to focus on the fundamental characteristics of fast/Alfv\'en conversion, stripped of unnecessary detail, it is therefore useful to freeze out the slow mode by adopting the gravitationally stratified cold MHD model $c\to0$. This provides a benchmark for fast-to-Alfv\'en mode conversion in more complex atmospheres. Assuming a uniform inclined magnetic field and an exponential Alfv\'en speed profile with density scale height $h$, the Alfv\'en conversion coefficient depends on three variables only; the dimensionless transverse-to-the-stratification wavenumber $\kappa=kh$, the magnetic field inclination from the stratification direction $\theta$, and the polarization angle $\phi$ of the wavevector relative to the plane containing the stratification and magnetic field directions. We present an extensive exploration of mode conversion in this parameter space and conclude that near-total conversion to outward-propagating Alfv\'en waves typically occurs for small $\theta$ and large $\phi$ ($80^\circ$--$90^\circ$), though it is absent entirely when $\theta$ is exactly zero (vertical field). For wavenumbers of helioseismic interest, the conversion region is broad enough to encompass the whole chromosphere.
\end{abstract}

\keywords{Sun, oscillations; sunspots; Magnetohydrodynamics (MHD)}


\section{Introduction}
Magneto\-hydro\-dynamic (MHD) linear mode conversion is an important process in solar active regions and in the overlying atmosphere. Fast-to-slow conversion is implicated in the absorption of p-modes by sunspots \citep{cbz,cccbd} and is well-understood in terms of local analysis around the Alfv\'en/acoustic equipartition level $a=c$ where the sound speed $c$ and Alfv\'en speed $a$ coincide \citep{sc06}. 

{To be specific, our primary target in this paper relates to mode coupling in sunspots between the well-known p-mode seismic wave field of the solar interior and oscillations of various types in the overlying atmosphere. Sunspot seismology is one of the most difficult issues confronting helioseismology today \citep{moradi10}. Early attempts at seismic inversion using time-distance helioseismology interpreted phase shifts in terms of temperature variations in the first few Mm below the surface, and obtained a two-layer thermal structure that is at odds with other inversions \citep[see Figure 19 of][]{moradi10}. Clearly, the near surface layers of sunspots are dominated by magnetic fields, with the plasma $\beta$ typically passing through unity a few hundred km below the surface in umbrae, and at about the surface in penumbrae. This has the effect of both mandating fast-to-slow mode conversion where the sound and Alfv\'en speeds are equal  \citep{sc06,cally07,cgd08}, and radically changing the phase of the fast wave that emerges through the surface and then reflects back downward to rejoin the subsurface helioseismic field (see \citealt{cally07}, Figure 3, and \citealt{cally09}). The combination of these two effects must confound any seismic inversion effort that attributes phase anomalies to sound speed perturbations alone. Conversion between the fast and slow magnetoacoustic waves may occur in two dimensions (2D), where magnetic field and the wavevector lie in a vertical plane ($x$-$z$ say), and also in 3D.}

{A further insufficiently modelled effect associated with this picture is that of fast-to-Alfv\'en conversion, occurring near the fast wave reflection point when the waves are directed at an angle to the magnetic plane (3D). Typically, this occurs a few hundred km above the $a=c$ level. Conversion was quantified for uniform inclined field in a specific simple solar atmospheric model by \citet{cg08} and confirmed in simulations by \citet{kc11}. The process has the potential to both remove energy from the reflecting fast wave and alter its phase before it re-enters the interior, with obvious implications for helioseismic inference, and for our interpretation of observations of waves in sunspot atmospheres. This explains our concentration on fast-to-Alfv\'en conversion. However, the reverse Alfv\'en-to-fast process is governed by the same considerations. It is well-known that the transition region can act as a powerful reflector of Alfv\'en waves \citep{hollweg81,cvb05}, so escaping Alfv\'en waves may in fact be partially `reabsorbed' by the fast wave field after such reflection, though presumably with radically different phase. This would further complicate the seismology of sunspots. Alfv\'en reflection and reabsorption is beyond the scope of the present study though.}

{In light of this context, it is appropriate to concentrate on the vertical rather than horizontal variations in Alfv\'en speed in the region of interest. The Alfv\'en scale height in a low sunspot atmosphere may be of the order of 100--200 km typically, whereas horizontal variations may take place over distances of order 10 Mm. In the interests of mathematical tractability, we therefore ignore the latter. However, full 3D numerical simulations in a realistic spreading sunspot magnetic field have very clearly confirmed the broad conclusions to be presented here (Khomenko \& Cally, in preparation).}

{Irrespective of these considerations, the basic process of fast-to-Alfv\'en conversion is of fundamental interest in MHD wave theory, and is pursued here in that spirit.}

\begin{figure}
\begin{center}
\includegraphics[width=0.5\hsize]{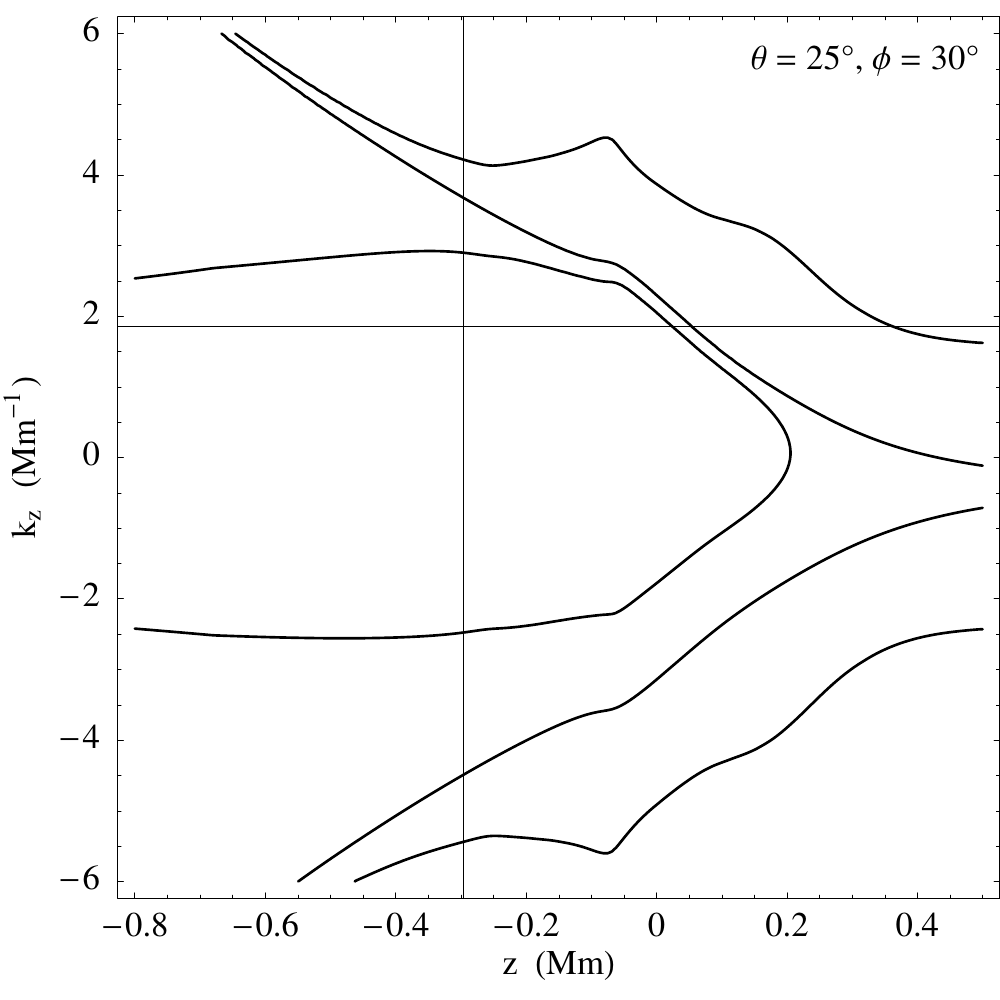}
\caption{Typical $z$-$k_z$ propagation diagram (where $z$ is height) for a 5 mHz wave in the model of Cally and Goossens (2008) where the magnetic field is inclined at angle $\theta=25^\circ$ from the vertical and the wavevector is oriented $\phi=30^\circ$ out of the vertical magnetic plane. The inner lobe represents the fast wave and the outer branches correspond to the slow wave. The intermediate curves are the Alfv\'en wave. The vertical line represents the equipartition depth at which $a=c$. Note that the fast wave reflects at about 200 km whilst the slow and Alfv\'en waves extend indefinitely upwards. The slow wave does reflect at lower frequencies below the ramp-reduced acoustic cutoff $\omega_c \cos\theta$, though that is not relevant to our discussion here.
}  \label{fig:prop}
\end{center}
\end{figure}

{The full details of specific} atmospheric models may be only weakly relevant to the strength of fast-to-Alfv\'en mode conversion, possibly implicated in the apparent ubiquity of Alfv\'en waves\footnote{There has been some controversy in the literature about whether these are truly Alfv\'en waves, or instead kink waves \citep{vanDNV}. Kink waves occur readily in media with transverse variations in the Alfv\'en speed, for example in atmospheres consisting of isolated or packed flux tubes, and propagate at a phase speed which is a weighted average of the internal and external Alfv\'en speeds. {The models we address here have no such transverse structure, and so are undoubtedly Alfv\'en waves. They may well be expected to couple to kink waves higher in the atmosphere though where coronal loop structure dominates.}} in the solar corona, as observed using the Solar Optical Telescope (SOT) aboard \emph{Hinode} \citep{deP07} and with the Coronal Multi-Channel Polarimeter (CoMP) at the National Solar Observatory, New Mexico \citep{tomczyk07}. Figure \ref{fig:prop} displays a typical example where fast-to-slow conversion occurs near the $a=c$ equipartition depth but the fast and Alfv\'en modes come to near-coincidence considerably higher, where $c\ll a$. This suggests that we seek to simplify the analysis of fast-to-Alfv\'en mode conversion by `freezing out' the slow wave. This can be done by examining a cold MHD plasma $c=0$ in which the uncoupled fast wave dispersion relation becomes simply $\omega^2=a^2 |\k|^2$ whilst the Alfv\'en wave remains $\omega^2=a^2\kpar^2$, where $\omega$ and $\k$ are the frequency and wavevector as usual, and the subscript $\scriptstyle\parallel$ indicates a component in the direction of the magnetic field . The cold MHD approximation corresponds to the $\beta=0$ limit, where $\beta$ is the ratio of gas to magnetic pressure. We briefly discuss the applicability of the cold-plasma approximation to solar atmospheric mode conversion in Section \ref{disc} by comparing with some relevant warm plasma results.

One advantage of the $\beta=0$ approximation is that we may ignore explicit gravity and just retain the density stratification it produces. In large scale magnetic structures such as sunspots the density scale height is typically much smaller than the scale of variation of magnetic field $\B_0$. We may assume $\B_0$ is uniform and therefore that the variation in the Alfv\'en speed $a=B_0/\sqrt{\mu\,\rho}$ is entirely due to the density stratification. Our aim in this paper is to quantify fast-to-Alfv\'en mode conversion for this simplified cold plasma model in the expectation that it will apply broadly to more complex structures and atmospheres.

\begin{figure}
\begin{center}
\includegraphics[width=0.7\hsize]{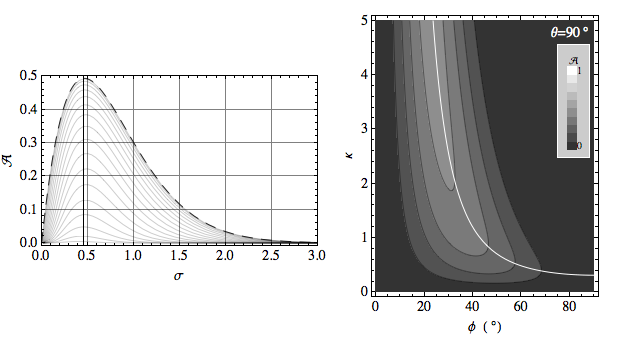}
\caption{Left panel: Absorption coefficient $\mathscr{A}$ as a function of $\sigma=(k h)^{2/3}\sin^2\phi$ for $a^2\propto e^{-x/h}$ and $\phi=5^\circ$, $10^\circ$, \ldots, $85^\circ$ (thin curves, top to bottom) for the resonant case $\theta=90^\circ$ \citep[from][]{ca10}. The heavy dashed curve represents the limit $\phi\to0^\circ$. The vertical line is at $\sigma=\sigma_0=0.4644$. Right panel: Contour plot of $\mathscr{A}$ against $\phi$ and $\kappa=k h$. The maximum in absorption occurs as $\kappa\to\infty$, $\phi\to0$ along the curve $\kappa^{2/3}\sin^2\phi=\sigma_0$ (delineated in white). }
\label{fig:Asig}
\end{center}
\end{figure}

This model has already been partially explored by \citet{ca10} and \citet{hc11}, who examined the situation where the magnetic field is oriented perpendicular to the direction of stratification $x$ and the density decreases exponentially with scale distance $h$, \emph{i.e.}, $a^2\propto \exp[x/h]$. In the sunspot model this would be horizontal magnetic field  and may be relevant to penumbra and canopy. It is a singular case in which resonant absorption occurs at the Alfv\'enic critical level where $\omega^2=a^2\kpar^2$. Nevertheless, \citeauthor{ca10} succeeded in interpreting this absorption as a mode conversion, thereby clarifying the relationship between these two related processes. With $y$ and $z$ wavenumbers $k_y$ and $k_z$ fixed, they tabulated and plotted the fast-to-Alfv\'en conversion coefficient $\mathscr{A}(\sigma,\phi)$, where $\tan\phi=k_y/k_z$ denotes the wave polarization, $\sigma=\kappa^{2/3}\sin^2\phi$, $h$ is the uniform density scale length, \emph{i.e.}, $a^2\propto \exp[x/h]$, and $\kappa=(k_y^2+k_z^2)^{1/2}h=(\kappa_y^2+\kappa_z^2)^{1/2}$ is the dimensionless transverse wavenumber. Remarkably, $\mathscr{A}$ is limited by a maximum value $A_0=0.4937$ attained at $\sigma=\sigma_0=0.4644$ in the limit $\phi\to0$ (see Figure \ref{fig:Asig}a). Note though that $\mathscr{A}=0$ in the 2D case $\phi=0$; Alfv\'en conversion \emph{requires} a 3D component $\kappa_y$ in the wavevector. The seemingly contradictory result that absorption is maximal in that limit in fact applies only if $\sigma$ is kept fixed at $\sigma_0$ and hence that $\kappa\to\infty$ simultaneously.

Our task here is to generalize this result to arbitrary magnetic field orientation $\theta$ from the $x$-direction. Without loss of generality we may assume that $\B_0$ lies in the $x$-$z$ plane but that the incident wave may cut across it ($k_y\ne0$). This is more straightforward than the perpendicular field case since there is no resonant singularity in the equations.

\section{Equations}

Consider a cold MHD plasma with monotonic increasing Alfv\'en speed $a(x)$ (due to monotonic decreasing density) and uniform magnetic field $\B_0=B_0(\cos\theta,0,\sin\theta)$ in cartesian coordinates $(x,y,z)$, with $0^\circ\leq\theta<90^\circ$. The orientation of the stratification is arbitrarily chosen to be in the $x$ rather than the $z$ direction for consistency with \citet{ca10}. In the linear approximation the plasma displacement $\bxi(x,y,z,t)=\bxi(x)\exp[\ri(k_yy+k_zz-\omega t)]$ where $\bxi(x)=\xi_x\e_x+\xi_y\e_y+\xi_z\e_z= \xi_\perp \e_\perp+\xi_y\e_y$ then satisfies
\begin{equation}
\left(\partial_\parallel^2+\frac{\omega^2}{a^2}\right)\bxi  =-\grad_{\!\text p}\chi\,,                                \label{basiceqn}
\end{equation}
generalizing equation (20) of \citet{ca10}. Here $\chi=\Div\bxi$ is the dilatation, the subscript `$\scriptstyle\parallel$' denotes the parallel direction $(\cos\theta,0,\sin\theta)$, `$\scriptstyle\perp$' indicates the direction $(\sin\theta,0,-\cos\theta)$ perpendicular to the field in the $x$-$z$ plane, and `p' refers to the component in the plane perpendicular to $\B_0$. Since there is no restoring force in the parallel direction it follows that $\xi_\parallel=0$, and hence $\bxi=\bxi_{\text p}$.

The corresponding wave-energy (Poynting) flux is
\begin{equation}
\F = \frac{1}{\mu}\re\left[\E_1^*\vcross\B_1\right]=F_0 \im\left[\chi\, \bxi^*+
(\bxi^*\vdot \,\,\partial_\parallel\bxi)\e_\parallel\right]
\end{equation}
where $F_0=\omega B_0^2/\mu$, and $\E_1=-\boldv\vcross\B_0$ and $\B_1$ are the perturbed electric and magnetic fields respectively. It may be verified directly (with the aid of the product rule) that $\Div\F=0$ by contracting Equation (\ref{basiceqn}) with $\bxi^*$ and taking the imaginary part. Since $\F$ is clearly independent of $y$ and $z$ it follows that the $x$-component $F_x$ is constant, as one would expect. This is distinct from the resonant case $\theta=90^\circ$ where $F_x$ is only piecewise constant, with a discontinuous drop to zero at the Alfv\'en resonance.


Eliminating $\chi$ from (\ref{basiceqn}) results in
\begin{equation}
\begin{aligned}
\left(\partial_\parallel^2+\partial_\perp^2+\frac{\omega^2}{a^2}\right)\xi_\perp
 &= -\ri\, k_y \partial_\perp\xi_y \\
\left(\partial_\parallel^2+\frac{\omega^2}{a^2}-k_y^2\right)\xi_y &= -\ri\, k_y \partial_\perp\xi_\perp\, ,                                
\end{aligned}   \label{parxi}
\end{equation}
or in terms of $x$-derivatives only {(for computational purposes)},
\begin{equation}
\begin{aligned}
\left(\partial_x^2+\frac{\omega^2}{a^2}-k_z^2\right)\xi_\perp
& =-\ri\, k_y (\sin\theta\,\partial_x-\ri\,k_z\cos\theta)\xi_y \\
\left((\cos\theta\,\partial_x+\ri\,k_z\sin\theta)^2+\frac{\omega^2}{a^2}-k_y^2\right)\xi_y & = -\ri\, k_y (\sin\theta\,\partial_x-\ri\,k_z\cos\theta)\xi_\perp\, .           
\end{aligned}      \label{odes}
\end{equation}
{In general, the fast and Alfv\'en waves are intricately intertwined in these equations. Their separate identities are more clearly seen in Equation \eqref{basiceqn}, where the left hand side exhibits the pure Alfv\'en operator and the right hand side represents the fast wave (characterized by the dilatation $\chi$) as a source term. The perturbation analysis of Section  \ref{pert} further expands on the coupling between the two wave types.}

We now specialize to the Alfv\'en profile defined by $\omega^2 h^2/a^2=\rme^{-x/h}$, where $h$ is the density scale length, and also define the dimensionless variables $s=\rme^{-x/h}$, $\X=\bxi/h$, $\kappa_y=k_yh=\kappa\sin\phi$, and $\kappa_z=k_zh=\kappa\cos\phi$. Note that fast mode reflection occurs (classically) {where $k_x=0$, \emph{i.e.}, at $\omega^2=a^2(k_y^2+k_z^2)$. In dimensionless form this is $s=\kappa^2$.  The Alfv\'en wave is well-described by the eikonal approximation where $s=\kappa_\parallel^2\gg1$.}

Defining $\U=(X_\perp,X_y,sX_\perp',s X_y')$, where the prime denotes the $s$ derivative, Equations (\ref{odes}) take the form
\begin{equation}
s\U'=\A\U    \label{mateqn}
\end{equation}
with
\begin{equation}
\A =
\begin{pmatrix}
\mathbf{0} & \I \\
\P & \Q
\end{pmatrix}    \label{A}
\end{equation}
where $\mathbf{0}$ and $\I$ are the $2\times2$ zero and identity matrices respectively, 
\begin{equation}
\P=
\begin{pmatrix}
 \kappa ^2 \cos ^2\phi -s & -\kappa ^2 \cos \theta  \cos
   \phi  \sin \phi  \\
 -\kappa ^2 \cos \phi  \sec \theta  \sin \phi  & \ \ \kappa^2(\tan^2\theta+\sin^2\phi) - s \sec ^2\theta 
   \end{pmatrix}
\end{equation}
and
\begin{equation}
\Q= \ri\, \kappa
\begin{pmatrix}
 0 &   \sin \theta  \sin
   \phi  \\ 
   \sec \theta  \sin \phi  \tan \theta  & \ 2 
   \cos \phi  \tan \theta 
\end{pmatrix}.
\end{equation}
Note that we may write $\A=\A_0+s\A_1$, where $\A_0$ and $\A_1$ are constant matrices.

Equation (\ref{mateqn}) may be solved numerically subject to the boundary conditions (i) the fast wave decays as $x\to+\infty$ ($s\to0^+$);
(ii) there is no incoming Alfv\'en wave at $x=+\infty$;
(iii) there is no incoming Alfv\'en wave at $x=-\infty$;
and (iv) the incoming fast wave at $x=-\infty$ carries unit $x$-flux. To apply these conditions in practice we develop a Frobenius expansion about $s=0$ and a WKB solution valid for large $s$ in the Appendix.

\section{Numerical Results}

\begin{figure}
\begin{center}
\includegraphics[width=0.5\hsize]{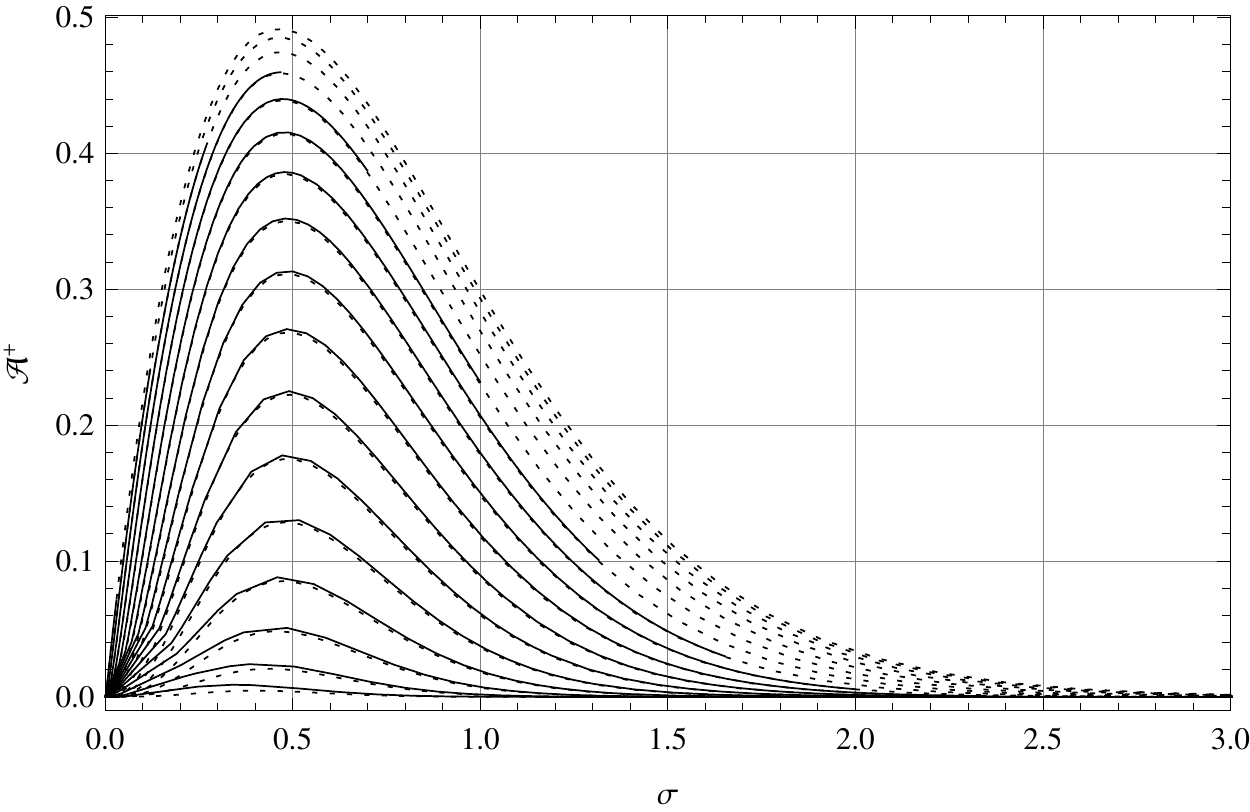}
\caption{Same as Figure \ref{fig:Asig}a, except for $\theta=85^\circ$ (full curves). For comparison, the Alfv\'en absorption curves of Figure \ref{fig:Asig} for the resonant case $\theta=90^\circ$ are overlaid as dotted curves. The full curves are truncated at $\kappa=8$ for numerical reasons, which is progressively more restrictive as $\phi$ increases.}
\label{fig:Acompare85}
\end{center}
\end{figure}

The resonant perpendicular field case $\theta=90^\circ$ discussed in \citet{ca10} is commonly thought to differ fundamentally from the non-resonant cases $0^\circ<\theta<90^\circ$ addressed here in that it is singular at the Alfv\'en resonance $\omega=a \kpar$, and the `Alfv\'en absorption' $\mathscr{A}$ is generally perceived as a local resonant absorption rather than a simple mode conversion to a transmitted wave. {(We have defined $\mathscr{A}$ as the fraction of the incident fast wave energy flux that is ultimately converted to an Alfv\'en wave. Fraction $1-\mathscr{A}$ belongs to the reflected fast wave.)} Nevertheless, Figure \ref{fig:Acompare85} for $\theta=85^\circ$ suggests that the two situations are not that different after all: the Alfv\'en absorption/conversion coefficients are near-identical. Clearly, despite the singular nature of the $\theta=90^\circ$ `resonant absorption' case, it is a continuous extension of the $\theta<90^\circ$ `mode conversion' cases. However, this is not to say that resonant absorption is illusory. See the discussion in Section 6 of \citet{ca10} for further detail, in particular regarding the significance of trivial Fourier-transformable directions.

\begin{figure*}
\begin{center}
\includegraphics[width=.73\hsize]{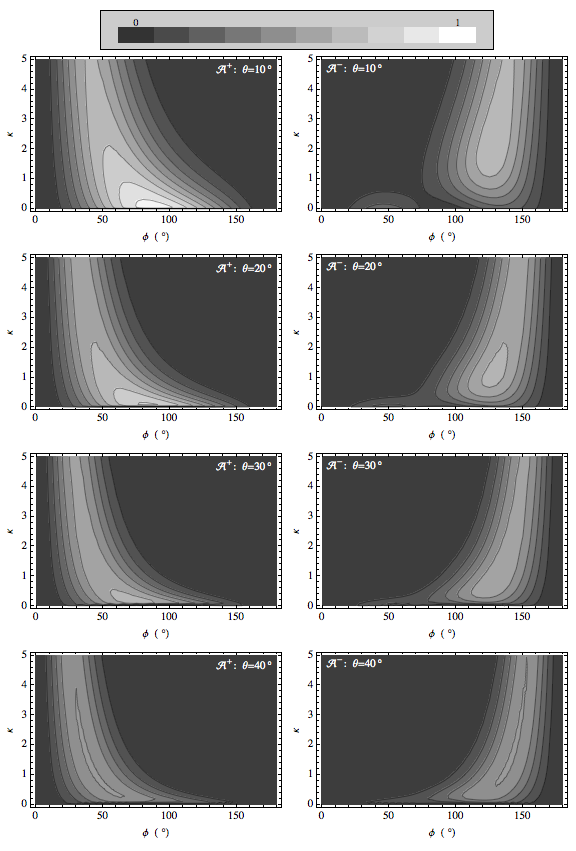}
\caption{Left column: The forward Alfv\'en conversion coefficient $\mathscr{A}^+$ as a function of $\phi$ and $\kappa$ for $\theta=10^\circ$, $20^\circ$, $30^\circ$, and $40^\circ$ (top to bottom) as labelled. Right column: The reverse Alfv\'en conversion coefficient $\mathscr{A}^-$ for the same cases. In all cases, the contours are 0.1, 0.2, \ldots, 0.9. The respective maxima in $\mathscr{A}^+$ are: 0.951 for $\theta=10^\circ$ ($\kappa=0.056$, $\phi=88.0^\circ$); 0.814 for $\theta=20^\circ$ ($\kappa=0.107$, $\phi=82.9^\circ$); 
0.647 for $\theta=30^\circ$ ($\kappa=0.170$, $\phi=73.9^\circ$); 
0.505 for $\theta=40^\circ$ ($\kappa=1.29$, $\phi=41.0^\circ$).
}
\label{fig:conts10to40}
\end{center}
\end{figure*}

\begin{figure*}
\begin{center}
\includegraphics[width=.73\hsize]{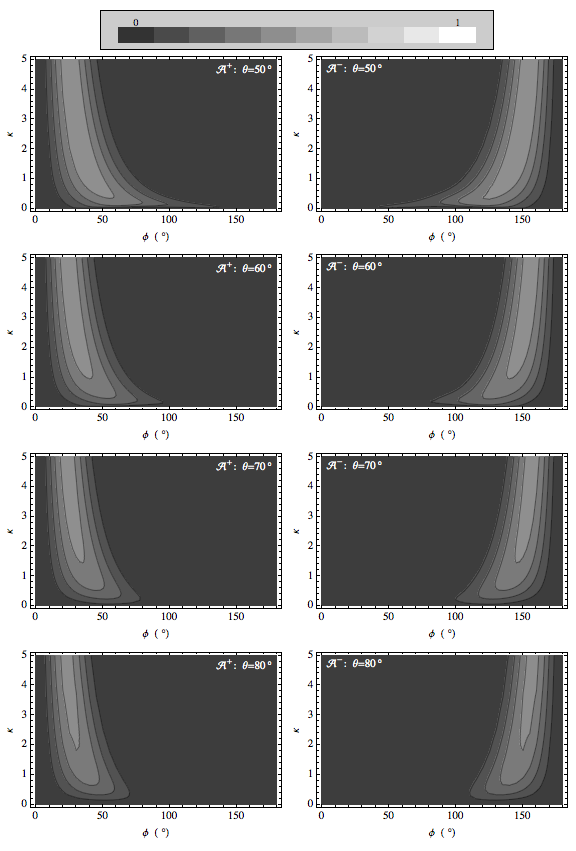}
\caption{Same as Figure \ref{fig:conts10to40}, but for $\theta=50^\circ$, $60^\circ$, $70^\circ$, and $80^\circ$ (top to bottom). The maxima in $\mathscr{A}^+$ occur above $\kappa=5$ in all cases.
}
\label{fig:conts50to80}
\end{center}
\end{figure*}

Figures \ref{fig:conts10to40} and  \ref{fig:conts50to80} show graphically how the forward Alfv\'en conversion coefficient $\mathscr{A}^+$ (conversion to the rightward propagating Alfv\'en wave) and reverse coefficient $\mathscr{A}^-$ (conversion to the leftward Alfv\'en wave) vary with transverse wavenumber $\kappa$ and wave polarization $\phi$ for magnetic field inclination $\theta=10^\circ$, $20^\circ$, \ldots, $80^\circ$.

Clearly, $\mathscr{A}^+$ is favoured at $\phi<90^\circ$ and $\mathscr{A}^-$ generally dominates on $\phi>90^\circ$. This is as expected, since maximal mode conversion is associated with an alignment of the phase velocities of the donor and recipient wave. If this occurs on the `upstroke' of the fast wave's path then $\mathscr{A}^+$ is stronger, but if it is on the `downstroke', after reflection, then $\mathscr{A}^-$ is the more significant. Since the approximate alignment occurs before the fast wave reflects for $0^\circ<\theta<90^\circ$ and $-90^\circ<\phi<90^\circ$, and after for $\phi>90^\circ$, the numerical results are plausible. The increasing symmetry between $\mathscr{A}^+$ and $\mathscr{A}^-$ as they weaken with increasing $\theta$ is also consistent with this interpretation.

Both $\mathscr{A}^\pm$ vanish in various limits, specifically (i) $\kappa=0$; (ii) $\theta=0^\circ$; and (iii) $\phi=0^\circ$ and $180^\circ$. The contour figures \ref{fig:conts10to40} and  \ref{fig:conts50to80} indicate though that the drop to zero in the $\kappa\to0$ limit is very sharp for small $\theta$. This is also seen in Table \ref{tab10} for $\mathscr{A}^+$ at $\theta=10^\circ$.

\begin{figure}
\begin{center}
%
%
\includegraphics[width=.8\hsize]{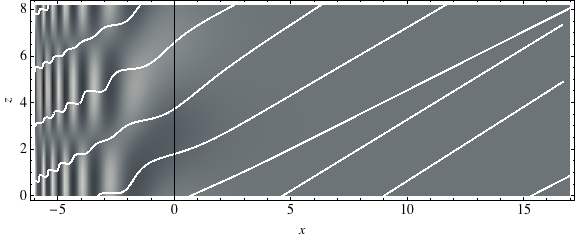}\\[6pt]
\includegraphics[width=.8\hsize]{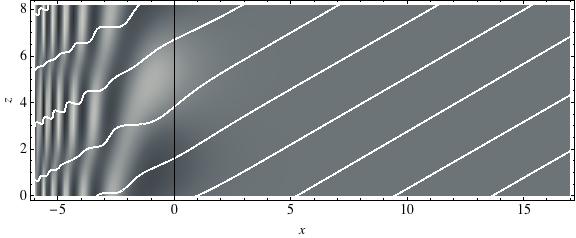}
\caption{Top: A greyscale version of a frame from a supplemental animation for the case $\kappa=1$, $\theta=30^\circ$, $\phi=40^\circ$. The shading represents $\chi=\Div\bxi$, and hence the fast wave, which classically reflects at $x=0$ (vertical black line). A selection of magnetic field lines are overplotted. In equilibrium they are all inclined at $30^\circ$ but when perturbed they oscillate up and down due to the Alfv\'en wave. Beyond about $x=0$ though they remain almost straight due to the increasingly long Alfv\'en wavelength. In the animations, the field line colours represent back-and-forth motion in the $y$ direction. Bottom: Same, but for $\phi=140^\circ$, exhibiting much less Alfv\'en conversion.}
%
%
%
\label{fig:still}
\end{center}
\end{figure}

Figure \ref{fig:still} (top) shows a single frame from an animation for $\kappa=1$, $\theta=30^\circ$, $\phi=40^\circ$ available as a supplement to this paper. It depicts the fast wave field through the green-yellow background shading representing $\chi$. Several magnetic field lines are overplotted, and wave back and forth in the animation. In the still frame shown here it is clear that to the right of the reflection point ($x=0$ in this case) the field lines are nearly straight, because of the very long Alfv\'enic wavelength, but tilted and moving with respect to each other. In contrast, the second supplemental animation (Figure \ref{fig:still} bottom) for the same case but with $\phi=140^\circ$ (exactly equivalent to reversing $\theta$ but keeping $\phi$ unchanged) exhibits very little Alfv\'enic action, as we may surmise from Figure \ref{fig:conts10to40}.

\begin{deluxetable}{llllllllllllllll} 
\tabletypesize{\footnotesize}
\rot
\tablewidth{0pt}\tablecolumns{16}
\tablecaption{Forward Alfv\'en conversion coefficient $\mathscr{A}^+$ for a range of $\kappa$ and $\phi$ at magnetic field inclination $\theta=10^\circ$. Similar tables for $\theta=20^\circ$, $30^\circ$, \ldots, $80^\circ$ are provided as supplementary material.\label{tab10}}
\tablehead{& \multicolumn{15}{c}{Polarization angle $\phi$}\\[4pt]
\colhead{$\kappa$} & \colhead{$5^\circ$} & \colhead{$15^\circ$} & \colhead{$25^\circ$} & \colhead{$35^\circ$} & \colhead{$45^\circ$} & \colhead{$55^\circ$} & \colhead{$65^\circ$} & \colhead{$75^\circ$} & \colhead{$85^\circ$}& \colhead{$95^\circ$} & \colhead{$105^\circ$} & \colhead{$115^\circ$} & \colhead{$125^\circ$} & \colhead{$135^\circ$} & \colhead{$145^\circ$} }
\startdata
0.000 & 0.000 & 0.000& 0.000& 0.000& 0.000& 0.000& 0.000& 0.000& 0.000& 0.000& 0.000& 0.000& 0.000& 0.000& 0.000\\
0.001 &  0.007 &  0.059 &  0.102 &  0.094 & 
   0.074 &  0.058 &  0.049 &  0.043 &  0.041 & 
   0.041 &  0.043 &  0.049 &  0.058 &  0.074 & 
   0.094 \\
 0.01 &  0.007 &  0.065 &  0.173 &  0.314 & 
   0.462 &  0.594 &  0.694 &  0.757 &  0.786 & 
   0.785 &  0.753 &  0.687 &  0.587 &  0.455 & 
   0.308 \\
0.1 & 0.008 & 0.070 & 0.185 & 0.338 & 0.509 & 0.673 & 0.811 & 0.905 & 0.945 & 0.927 & 0.855 & 0.739 & 0.593 & 0.435 & 0.282 \\
 0.2 & 0.008 & 0.075 & 0.196 & 0.356 & 0.529 & 0.692 & 0.821 & 0.901 & 0.924 & 0.889 & 0.804 & 0.681 & 0.537 & 0.387 & 0.248 \\
 0.3 & 0.009 & 0.079 & 0.207 & 0.372 & 0.547 & 0.706 & 0.826 & 0.892 & 0.898 & 0.847 & 0.751 & 0.624 & 0.482 & 0.342 & 0.216 \\
 0.4 & 0.009 & 0.083 & 0.216 & 0.386 & 0.563 & 0.717 & 0.827 & 0.878 & 0.868 & 0.804 & 0.698 & 0.569 & 0.432 & 0.301 & 0.187 \\
 0.5 & 0.010 & 0.087 & 0.225 & 0.399 & 0.576 & 0.726 & 0.825 & 0.862 & 0.837 & 0.760 & 0.647 & 0.517 & 0.385 & 0.263 & 0.161 \\
 0.6 & 0.010 & 0.090 & 0.234 & 0.411 & 0.588 & 0.732 & 0.821 & 0.844 & 0.805 & 0.717 & 0.598 & 0.468 & 0.342 & 0.230 & 0.138 \\
 0.7 & 0.011 & 0.094 & 0.242 & 0.423 & 0.599 & 0.737 & 0.815 & 0.824 & 0.772 & 0.675 & 0.552 & 0.423 & 0.303 & 0.200 & 0.119 \\
 0.8 & 0.011 & 0.097 & 0.249 & 0.433 & 0.608 & 0.740 & 0.807 & 0.804 & 0.739 & 0.634 & 0.508 & 0.382 & 0.268 & 0.174 & 0.101 \\
 0.9 & 0.012 & 0.101 & 0.257 & 0.443 & 0.616 & 0.742 & 0.798 & 0.782 & 0.707 & 0.594 & 0.467 & 0.344 & 0.237 & 0.151 & 0.087 \\
 1.0 & 0.012 & 0.104 & 0.264 & 0.452 & 0.624 & 0.742 & 0.788 & 0.760 & 0.675 & 0.557 & 0.429 & 0.309 & 0.208 & 0.130 & 0.074 \\
 1.5 & 0.014 & 0.118 & 0.295 & 0.490 & 0.649 & 0.732 & 0.728 & 0.650 & 0.529 & 0.396 & 0.275 & 0.179 & 0.109 & 0.062 & 0.032 \\
 2.0 & 0.016 & 0.131 & 0.321 & 0.519 & 0.660 & 0.708 & 0.660 & 0.546 & 0.407 & 0.277 & 0.174 & 0.102 & 0.056 & 0.029 & 0.014 \\
 2.5 & 0.017 & 0.143 & 0.344 & 0.541 & 0.663 & 0.676 & 0.592 & 0.455 & 0.311 & 0.192 & 0.109 & 0.057 & 0.028 & 0.013 & 0.006 \\
 3.0 & 0.018 & 0.154 & 0.364 & 0.558 & 0.659 & 0.640 & 0.527 & 0.376 & 0.236 & 0.132 & 0.068 & 0.032 & 0.014 & 0.006 & 0.002 \\
 3.5 & 0.020 & 0.164 & 0.382 & 0.571 & 0.650 & 0.602 & 0.467 & 0.309 & 0.178 & 0.091 & 0.042 & 0.018 & 0.007 & 0.003 & 0.001 \\
 4.0 & 0.021 & 0.173 & 0.397 & 0.580 & 0.638 & 0.563 & 0.411 & 0.253 & 0.134 & 0.062 & 0.026 & 0.010 & 0.003 & 0.001 & 0.000 \\
 4.5 & 0.022 & 0.182 & 0.412 & 0.587 & 0.624 & 0.525 & 0.361 & 0.207 & 0.101 & 0.043 & 0.016 & 0.005 & 0.002 & 0.001 & 0.000 \\
 5.0 & 0.024 & 0.190 & 0.424 & 0.591 & 0.607 & 0.488 & 0.316 & 0.169 & 0.076 & 0.029 & 0.010 & 0.003 & 0.001 & 0.000 & 0.000
\enddata 
\end{deluxetable}

\begin{figure}
\begin{center}
\includegraphics[width=0.5\hsize]{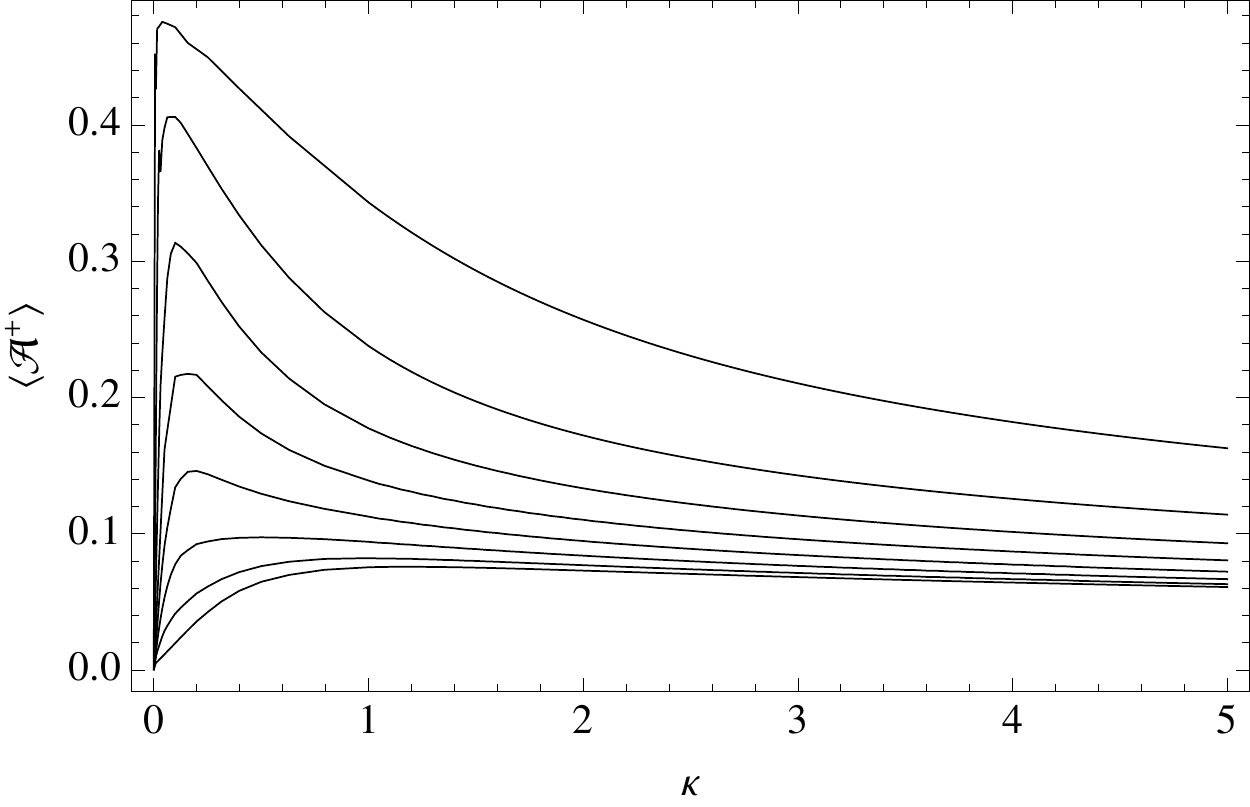}
\caption{$\mathscr{A}^+$ averaged over all $\phi$. The various curves are for $\theta=10^\circ$, $20^\circ$, \ldots, $80^\circ$ (top to bottom).}
\label{fig:Aav}
\end{center}
\end{figure}

Recognizing that wave direction $\phi$ is likely to be uniformly distributed in most circumstances, Figure \ref{fig:Aav} plots the $\phi$-averaged Alfv\'en conversion coefficient $\langle\mathscr{A}^+\rangle$ against $\kappa$ for a range of field inclinations $\theta$.
 
\section{Failure of Local Analysis}  \label{local}
\begin{figure}
\begin{center}
\includegraphics[width=0.4\hsize]{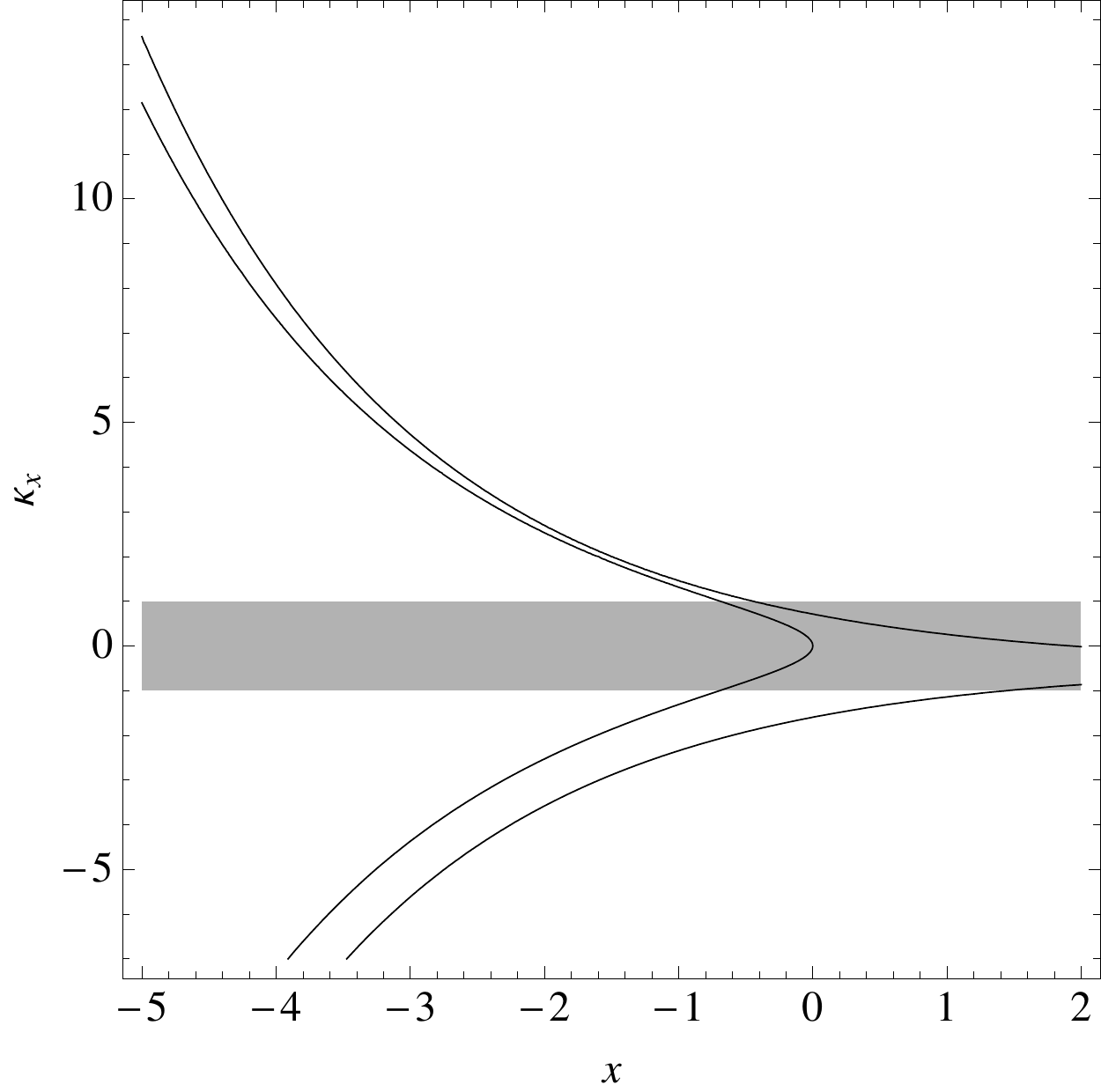}
\caption{The $x$-$\kappa_x$ phase plane for $\kappa=1$, $\theta=30^\circ$, $\phi=40^\circ$. The inner lobe represents the fast wave and the two outer wings are the Alfv\'en wave. The grey shading corresponds to $|\kappa_x|<1$. Formally the eikonal approximation is valid only on $|\kappa_x|\gg1$, though the dispersion curves shown here are still instructive at small $\kappa_x$, correctly indicating reflection of the fast wave at around $x=0$ and continued propagation of the Alfven wave as $x\to+\infty$.}
\label{fig:phaseplane}
\end{center}
\end{figure}

Mode conversion between fast and slow magnetoacoustic waves is amenable to local analysis \citep{sc06} using the method of \citet{tkb} or related WKB-based techniques since the conversion region is characterized by an avoided crossing of generic saddle point topology in the appropriate phase space. Tests against exact solutions in a gravitationally stratified isothermal atmosphere suggest very satisfactory accuracy when the gap between modes is small to moderate \citep{hc09}. As already indicated in Figure \ref{fig:prop} though, fast/Alfv\'en interactions typically involve a long and definitely non-local conversion region which may extend for several scale heights and which does not display the required saddle structure. Figure \ref{fig:phaseplane} illustrates the phase structure for the cold plasma case at hand, again displaying the long distributed interaction between the fast and Alfv\'en waves.  The dispersion relation $\mathcal{D}=(s-|\bkappa|^2)(s-\kappa_\parallel^2)=0$ clearly exhibits the two apparently disjoint modes, which in $s$--$\kappa_x$ space are simply non-intersecting parabolae. Attempts at a local analysis around the point of closest approach of the two branches have produced unsatisfactory results. It does not appear that any such local analysis can adequately capture fast-to-Alfv\'en mode conversion in this scenario. Indeed, surprisingly, we shall see in Section \ref{pert} that the bulk of the mode conversion actually occurs around and beyond the fast wave reflection point rather than where the fast and Alfv\'en loci approach most closely.

\section{Perturbation Analysis}  \label{pert}
Although the numerical analysis is essentially complete, it is of interest to develop a perturbation solution that illustrates the nature and locality of the fast-to-Alfv\'en interaction at the core of the conversion process. We perturb about the purely fast case $\kappa_y=0$, $\xi_y=0$. Equation (\ref{basiceqn}) is rewritten
\begin{equation}
\begin{aligned}
(\partial_x^2-\kappa_z^2+e^{-x})\xi_\perp &= -i\,\kappa_y\,\partial_\perp\xi_y \\[4pt]
(\partial_\parallel^2+e^{-x})\xi_y &= -i\,\kappa_y\,\partial_\perp\xi_\perp +\kappa_y^2\xi_y\, .
\end{aligned}
\end{equation}
The fully reflective fast wave with $\kappa_y=0$ and $\kappa_z=\kappa$ has solution
\begin{equation}
\xi_{\perp0} = J_{2\kappa}(2\,e^{-x/2}) 
= \half\left(
H_{2\kappa}^{(1)}(2\,e^{-x/2})  + H_{2\kappa}^{(2)}(2\,e^{-x/2})
\right) =\xi_{\perp0}^- + \xi_{\perp0}^+  \label{xiperp0}
\end{equation}
as a Bessel function of the first kind (standing wave), or alternatively in terms of leftward and rightward propagating Hankel functions. The incident $x$-component of wave energy flux associated with $\xi_{\perp0}^+$ is simply $F_\text{in} =F_0/4\pi$.

\begin{figure}
\begin{center}
\includegraphics[width=0.5\hsize]{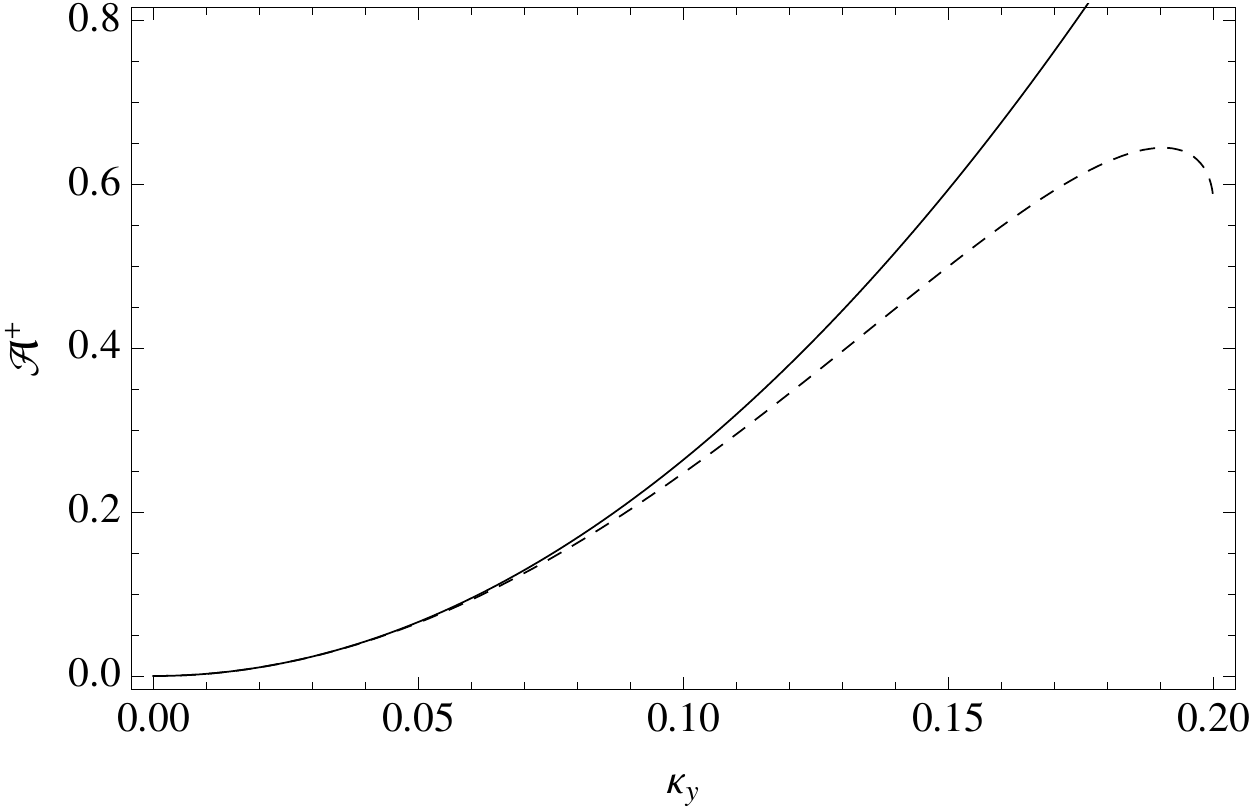}
\caption{Perturbation theory Alfv\'en conversion coefficient $\mathscr{A}^+$ (full curve) given by Equation \eqref{Apert} as a quadratic function of $\kappa_y$ for $\kappa=0.2$, $\theta=30^\circ$. The exact numerical result is shown dashed. For this case $B(\infty)=-0.6539-1.257\, i$.}
\label{fig:ApertComp}
\end{center}
\end{figure}

The transverse displacement $\xi_y$ appears at first order in $\kappa_y$. The correction $\xi_{\perp2}$ to $\xi_\perp$ is of second order and will not be required. At first order
\begin{equation}
\xi_{y1} \sim  -\frac{\kappa_y\pi}{2}\,e^{-i\kappa_z x\tan\theta}\sec^2\theta
\left\{
A(x)\,H_0^{(1)}(2e^{-x/2}\sec\theta) + B(x)\,H_0^{(2)}(2e^{-x/2}\sec\theta)
\right\}\, ,
\end{equation}
where
\begin{equation}
\begin{aligned}
A(x) &= \int_x^\infty e^{i\kappa_z X\tan\theta}H_0^{(2)}(2e^{-X/2}\sec\theta)\,\partial_\perp\xi_{\perp0}(X)\,dX\, ,\\[4pt]
B(x) &= \int_{-\infty}^x e^{i\kappa_z X\tan\theta}H_0^{(1)}(2e^{-X/2}\sec\theta)\,\partial_\perp\xi_{\perp0}(X)\,dX\, .
\end{aligned}  \label{AB}
\end{equation}
The limits on the integrals have been chosen so that there are no incoming Alfv\'en waves at either end. For large $x$ well beyond the interaction region, the upper limit of the $B$ integral may be replaced by $\infty$, whence
\begin{equation}
\xi_{y1}^+ \sim -\frac{\kappa_y\pi}{2} \sec^2\theta\, B(\infty)\,e^{-i\kappa_z x\tan\theta} H_0^{(2)}(2e^{-x/2}\sec\theta) \quad\text{as $x\to\infty$}, 
\end{equation}
carrying $x$-component of flux
\begin{equation}
F^+ \sim F_0 \im[
\xi_y^*\partial_\parallel\xi_y\cos\theta] =F_0 \frac{\kappa_y^2\pi}{4}\,|B(\infty)|^2\sec^2\theta\, .
\end{equation}
Hence
\begin{equation}
\mathscr{A}^+ = \kappa_y^2\pi^2\sec^2\theta\,|B(\infty)|^2
+ \mathscr{O}(\kappa_y^4).   \label{Apert}
\end{equation}
Figure \ref{fig:ApertComp} illustrates how well this quadratic result compares with the full numerical solution for small $\kappa_y$.

\begin{figure}
\begin{center}
\hfil\includegraphics[width=0.45\hsize]{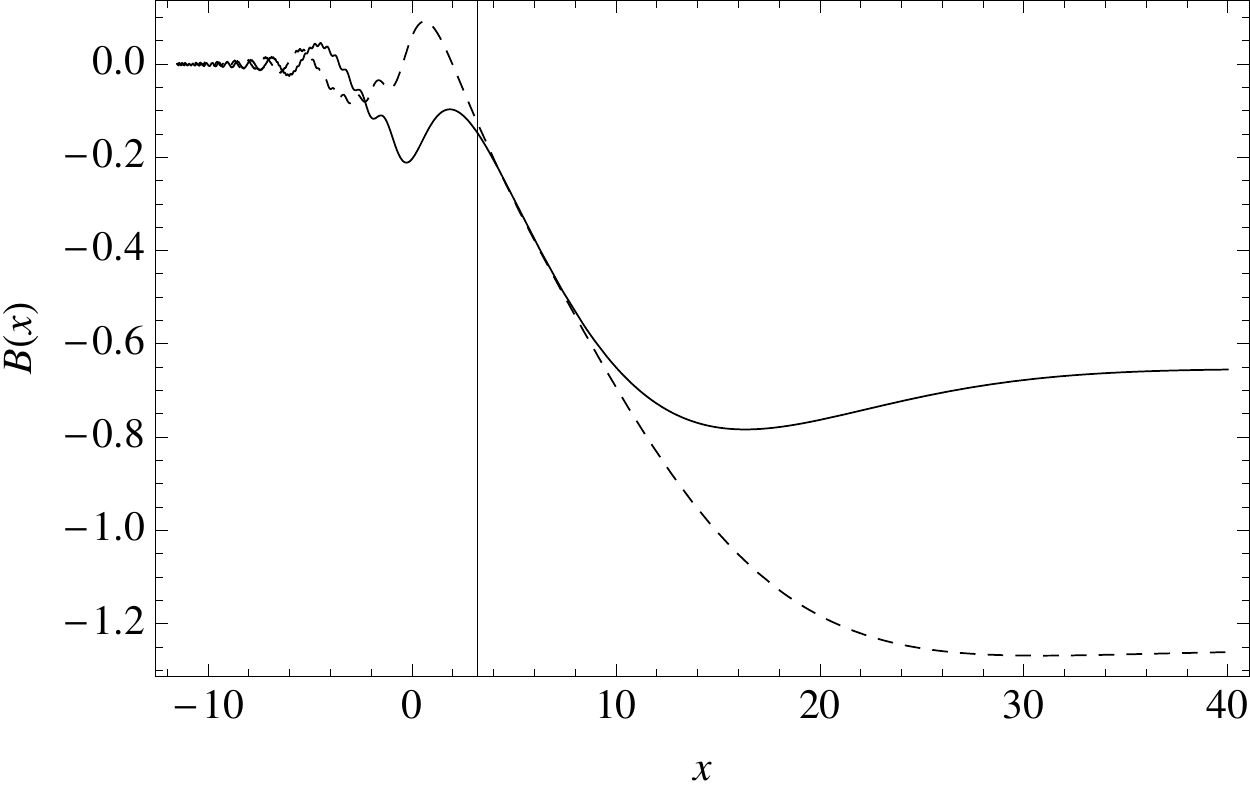}\hfil
\includegraphics[width=0.45\hsize]{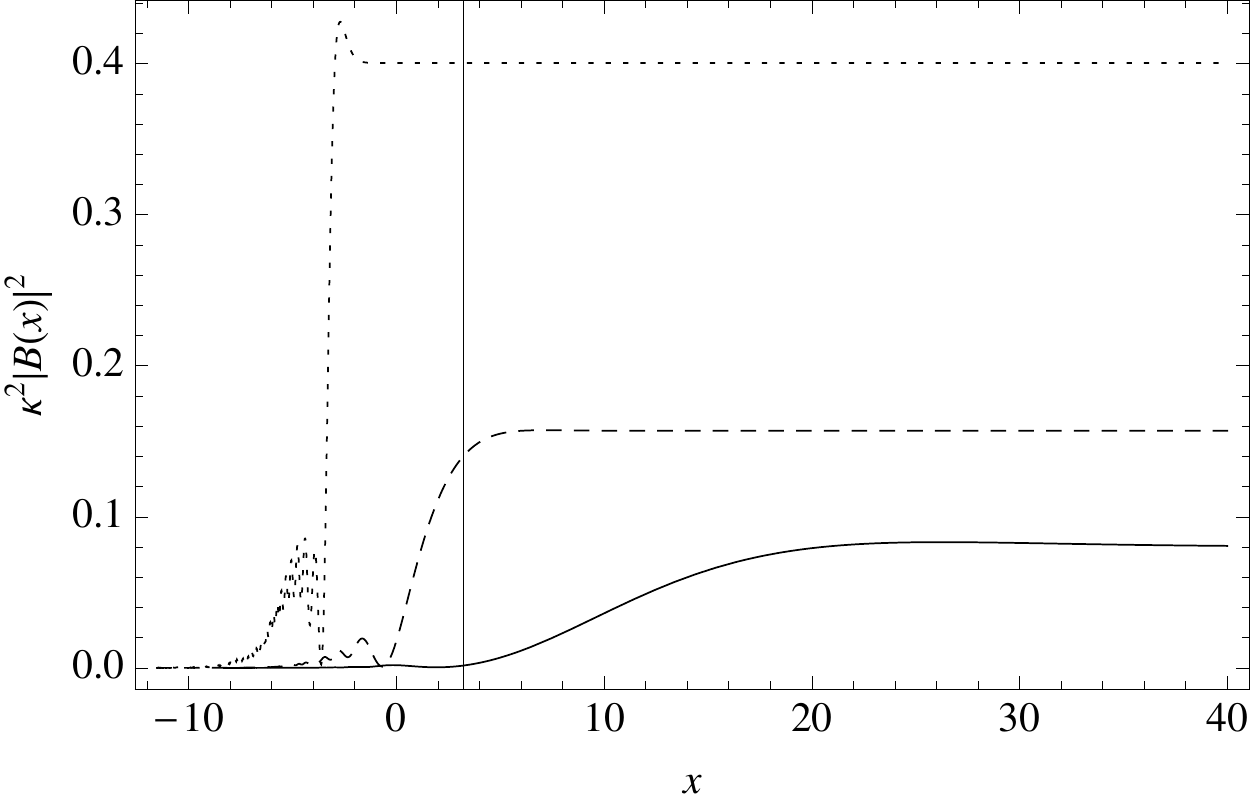}\hfil
\caption{Left: Interaction integral $B(x)$ for $\kappa=0.2$, $\theta=30^\circ$. The real and imaginary parts are shown as full and dashed curves respectively. Right: $\kappa^2|B(x)|^2$ for $\kappa=0.2$ (full curve), $\kappa=1$ (dashed), and $\kappa=5$ (dotted). The vertical line indicates the position of fast wave reflection in the $\kappa=0.2$ case. Reflection is at $x=0$ for $\kappa=1$ and $x=-3.22$ for $\kappa=5$. }
\label{fig:B}
\end{center}
\end{figure}

The interaction integral $B(x)$ defined in Equation \eqref{AB} provides a convenient picture of where fast-to-Alfv\'en mode conversion occurs. The jump in $|B|^2$ determines the amount of mode conversion, and the argument of $B(\infty)$ represents a phase shift.\footnote{The integrand of $B$ is extremely oscillatory and hence there is a severe loss of precision in integrating it numerically. In calculating $B(\infty)$ it is advisable to deform the integration contour $(-\infty,\infty)$ some distance (${}<2\pi$ to remain on the correct Riemann sheet) below the real line in the complex plane, where the oscillations are suppressed. This is done in calculating the $B(\infty)$ used in Figure \ref{fig:ApertComp} but, for purposes of illustration, Figure \ref{fig:B} uses integration on the real line.} Figure \ref{fig:B}, for the case $\kappa=0.2$, $\theta=30^\circ$, shows that it is in fact far more spread out in $x$ than might be expected from a dispersion diagram, with the major contribution occurring around and beyond the fast wave reflection point, where the eikonal approximation breaks down. This is consistent with the findings of \cite{ca10} for the $\theta=90^\circ$ case. At higher $\kappa$ though, as one might expect, the growth in $|B|^2$ is sharper and therefore the process more localized, but these higher wavenumbers are of lesser helioseismic interest.

Figure \ref{fig:B}b indicates that mode conversion is progressively more localized as $\kappa$ increases, as we might expect, though in all cases it predominantly occurs beyond the fast wave reflection point. However, since $\kappa\lesssim0.2$ typically for oscillations relevant to local helioseismology,\footnote{Assuming a density scale height of 150 km, the helioseimic degree $\ell=4640\,\kappa$.} it is clear that fast-to-Alfv\'en conversion is spread out over many scale heights above the reflection point for these waves, easily encompassing the whole chromosphere. Although the present model does not contain a transition region (TR) or corona, we may surmise that the jump in temperature of order 100 across the TR will sharply turn off this interaction. This is because the scale height $h$ and hence $\kappa=kh$ are similarly increased by two orders of magnitude.

\section{Comparison with $\beta>0$ Model}  \label{disc}

\begin{figure}
\begin{center}
\includegraphics[width=.5\hsize]{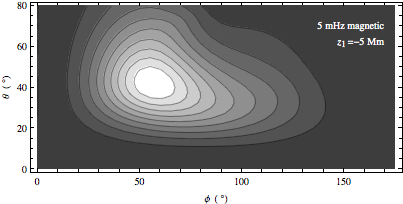}
\caption{The lower panel of Figure 2 of \citet{cg08}, showing the total magnetic (Alfv\'en) flux (absolute, not relative to an incident flux) at the top of a simplified (warm) solar-like atmospheric model threaded by a 2 kG uniform magnetic field and topped by an isothermal slab above $z=0.5$ Mm. The nine contours are equally spaced in flux. The bottom of the acoustic cavity hosting the waves in question is at $z_1=-5$ Mm, resulting in a horizontal wavenumber $k=1.37$ $\text{Mm}^{-1}$ for 5 mHz oscillations. This corresponds roughly to $\kappa=0.2$ in our dimensionless units.}
\label{fig:CG08mag}
\end{center}
\end{figure}

The aim in this paper is to fully characterize fast-to-Alfven mode conversion in a cold MHD plasma with uniform inclined magnetic field and an exponential Alfv\'en speed profile, neglecting slow-wave (acoustic) effects. However, it is instructive to compare our results with those of \citet{cg08} for a similarly configured warm plasma solar atmosphere to judge the extent to which the cold plasma model pertains to more realistic atmospheres. Figure \ref{fig:CG08mag} shows the magnetic wave-energy flux emerging at the top of this model for a case roughly comparable to our $\kappa=0.2$. This is absolute flux (not relative as we have focussed on here) resulting from a driving plane at $z_b=-4$ Mm and normalized by the total acoustic energy in $z_b<z<0$. Nevertheless, we might expect this figure to broadly coincide with the cold-plasma results.

\begin{figure}
\begin{center}
\includegraphics[width=.5\hsize]{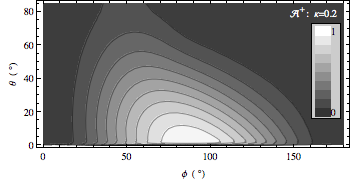}
\caption{$\mathscr{A}^+$ against $\phi$ and $\theta$ for fixed $\kappa=0.2$, roughly corresponding to Figure \ref{fig:CG08mag}. As usual, the contours are 0.1, 0.2, \ldots, 0.9. Note that $\mathscr{A}^+=0$ on $\theta=0$.}
\label{fig:kappaPt2}
\end{center}
\end{figure}

It is therefore surprising to view Figure \ref{fig:kappaPt2}, where strong Alfv\'en conversion persists almost all the way down to $\theta=0^\circ$, in stark contrast to the warm-plasma case where it is strongly suppressed for $\theta\lesssim15^\circ$. Despite this, the behaviours at larger $\theta$ are comfortingly similar, with strongest response around $\phi=50^\circ$--$70^\circ$ in both cases.

\begin{figure*}
\begin{center}
\includegraphics[width=\hsize]{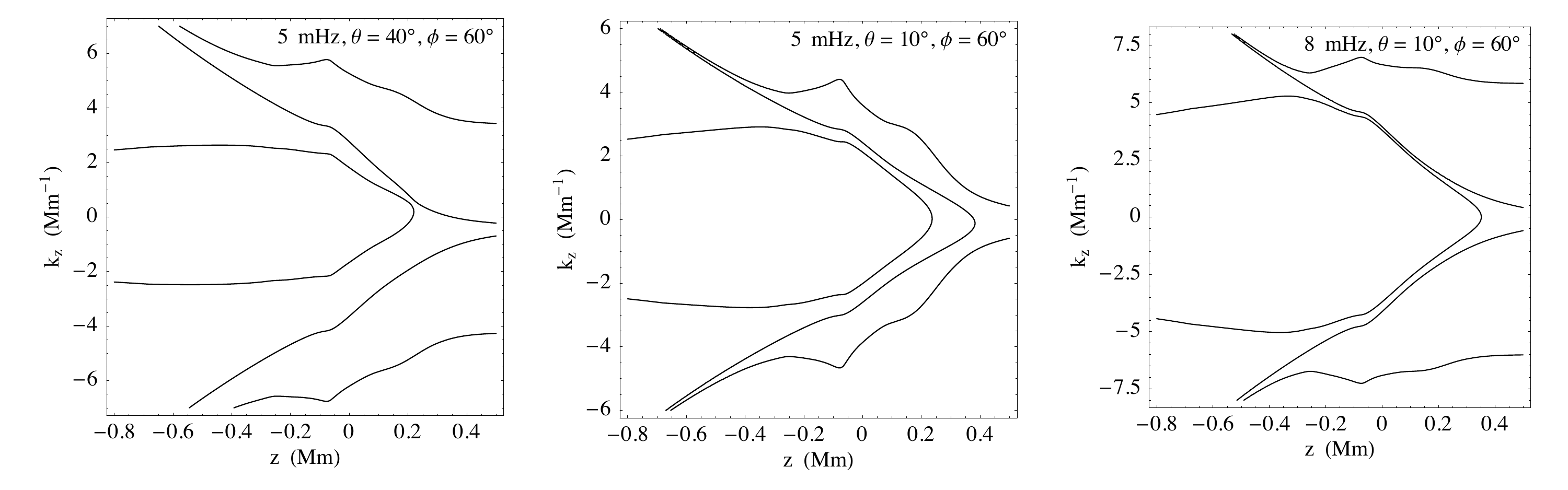}
\caption{Warm-plasma dispersion curves corresponding to $\theta=40^\circ$ (left) and $\theta=10^\circ$ (middle) with $\phi=60^\circ$ for the 5 mHz case of Figure \ref{fig:CG08mag}.  The inner lobes represent the fast wave, the outer wings the slow wave, and the intermediate branches the Alfv\'en wave. The right panel corresponds to higher frequency (8 mHz), again with $\theta=10^\circ$ and $\phi=60^\circ$.}
\label{fig:GONGdisp}
\end{center}
\end{figure*}

The reason for the disparity is made clear by the 5 mHz warm-plasma dispersion curves\footnote{The dispersion function used here is that derived by \citet{nc10},\emph{viz.},
\begin{equation*}
\calD=\omega^2 \omega_{\rm c}^2 a^2k_{\rm h}^2\sin^2\theta\sin^2\phi  +(\omega^2-a^2\kpar^2)\left[\omega^4-(a^2+c^2)\omega^2 k^2+a^2c^2k^2\kpar^2 + c^2N^2 k_{\rm h}^2-(\omega^2-a^2k^2\cos^2\theta) \omega_{\rm c}^2\right], 
\end{equation*}
where $\omega_\text{c}$ is the acoustic cutoff frequency, $N$ is the Brunt-V\"ais\"al\"a frequency, $a$ and $c$ are the Alfv\'en and sound speeds, $k=|\k|$ is the total wavenumber, and $k_h$ is the horizontal wavenumber. The first (additive) term on the right hand side breaks the separability of the Alfv\'en (\emph{i.e.}, $\omega^2-a^2\kpar^2$) and magnetoacoustic (expression in square brackets) modes, but it vanishes in the cold plasma regime where $\omega_\text{c}=0$.}  presented in Figure \ref{fig:GONGdisp}. At $\theta=40^\circ$, the fast-Alfv\'en interaction is much as in the cold plasma model. The slow wave locus is sufficiently far away as to have no effect. However, at $\theta=10^\circ$ the slow locus impinges on the Alfv\'en branch and actually causes it to turn over, suggesting reflection. The low-$\theta$ truncation in Figure \ref{fig:CG08mag} therefore makes sense; there will have been strong fast-to-Alfv\'en conversion as in the cold plasma case, but it quickly reflects with probably some weak Alfv\'en-to-slow (\emph{i.e.}, acoustic) conversion accompanying it. At lower frequencies, below $\omega_c\cos\theta$, the slow wave locus also turns over due to the acoustic cutoff effect. However, at higher frequencies (right panel) the generic cold plasma scenario is again apparent, even at small $\theta$.

\goodbreak
\section{Conclusion}  \label{concs}
This paper addresses the fundamental issue of mode conversion between fast magnetoacoustic waves and Alfv\'en waves in a cold stratified atmosphere. This is not to claim that the cold plasma approximation necessarily provides a full description of the solar atmosphere in conversion regions. It does though focus our attention solely on the conversion process. As indicated in Figure \ref{fig:GONGdisp} warm plasma effects can further intrude to modify the ultimate Alfv\'enic transmission, and these must be taken into account in a full description of wave propagation through these regions. Nevertheless, they are distinct processes, and warrant individual attention. For example, at small field inclination there may be very significant Alfv\'en conversion, but the Alfv\'en waves may subsequently be reflected back downward. This complex array of distinct but spatially adjacent processes has implications for the wave field in the solar atmosphere overlying active regions, and also for the helioseismic wave field beneath them. 

Although we have exclusively focussed on fast-to-Alfv\'en conversion, the reverse process may also be relevant in certain localized instances in the solar atmosphere. As a reversible process, conversion coefficients either way must be the same.

For the cold plasma at least, reference to Figure \ref{fig:Aav} indicates that a randomly oriented incident fast wave field with $\kappa\lesssim1$ encountering a moderately inclined magnetic region, such as is found in sunspot umbrae, might be expected to lose of the order of 20\% or more of its energy to Alfv\'en waves. Given the strong $\phi$ dependence though, we may also conclude that such inclined field can act as a powerful directional filter on the waves entering the solar atmosphere through active regions, suggesting possible observational tests.

In summary, the major insights we have gained from this study include:
\begin{enumerate}
\item Conversion to Alfv\'en waves is very sensitive to the direction $\phi$ of the incident fast wave, as illustrated in Figures \ref{fig:conts10to40} and \ref{fig:conts50to80}. 
\item Unlike fast-to-slow conversion, the interaction between fast and Alfv\'en waves is significantly spread across many scale heights. The major contribution to the conversion occurs around and beyond the fast wave reflection point rather than in the neighbourhood of the closest approach of the respective loci in phase space.
\item For transverse wavenumbers $\kappa=kh\lesssim0.2$ typical of local helioseismic waves, the conversion region is sufficiently spread to fill the whole chromosphere.
\item Although fast-to-Alfv\'en mode conversion identically vanishes for exactly vertical magnetic field ($\theta=0$) it can be near-total for small $\theta$ and small horizontal wavenumber $\kappa$, typically for $\phi\approx90^\circ$.
\item Nevertheless, warm-plasma effects conspire to reflect these Alfv\'en waves back downwards at small $\theta$.
\item At larger field inclinations ($\theta\gtrsim15^\circ$) conversion can still be significant, though at larger $\kappa$. These Alfv\'en waves are not reflected by warm-plasma effects.
\end{enumerate}

{The role of this study has been to illuminate the basic processes and dependences on wavenumber, magnetic field inclination, and wave attack direction of fast-to-Alfv\'en mode conversion. Subsequent work will build these insights into more elaborate numerical simulations of sunspots. Currently, all seismic and convective numerical sunspot models unrealistically limit the Alfv\'en speed in the overlying atmosphere for numerical reasons \citep[\emph{e.g.,}][]{han08,cgd08,rsk09}. The results presented here have implications for such models, as well as for our understanding of atmospheric waves. In particular, we postulate that the returning fast wave is important in understanding seismic travel time data, and therefore needs to me modelled correctly. The conversion process may also contribute to the recently-observed coronal transverse oscillations. In this scenario, the full width of active region chromospheres may be thought of as source region for these coronal waves, rather than (or in addition to) direct photospheric motions.}

{Finally, it is of interest to reflect on the recent modelling of coupling between kink and Alfv\'en waves in simple loop models with transverse Alfv\'en speed gradients by \cite{PWdeM10,PWdeM11}. A simple straight `loop' consisting of a uniform core and linear decreasing-density shoulders is shaken at its footpoint setting up an `Alfv\'en' wave in the core that rapidly decays by resonant absorption to leave oscillations restricted to the shoulders \cite[see][Figure 5]{PWdeM10}. However, such back-and-forth footpoint shaking preferentially initiates $m=1$ disturbances in the cylinder. Strictly, only $m=0$ (torsional) motions are genuine incompressive Alfv\'en waves in the cylindrical context, and these do not leak into the resonance as they cannot transport energy across field lines. Rather, the $m=1$ `Alfv\'enic' oscillations in the cores are fast waves, and their absorption at the resonance is in essence the process discussed in \cite{ca10} for the case of magnetic field transverse to the direction of inhomogeneity. The fact that \citeauthor{PWdeM10} see essentially total absorption rather than the 50\% maximum found by \citeauthor{ca10} results from multiple absorptions as the fast wave bounces back-and-forth across the tube. This is recognized in the schematic Figure 2 of \cite{PWdeM11}. In a zero $\beta$ plasma the core fast waves are the so-called (but unfortunately named) compressional Alfv\'en waves \citep[Section 4.3.2]{priest}.}

\appendix
\section{Details of Numerical Solution Method}
Analytic expressions for the fast and Alfv\'en solutions are required as $x\to\pm\infty$ in order to provide boundary conditions for numerical solution of the differential equations and to disentangle the two fast and two Alfv\'en modes.

\subsection{WKB at Large $s$}
A WKB solution is appropriate on $s=\rme^{-x/h}\gg1+\kappa^2$ since there $k_xh\gg1$. Substitute the eikonal ansatz $\U=\rme^{\ri\varphi}\sum_{n=0}^\infty\V_n$ into $s\U'=\A\U$, assuming $\V_0\ll\V_1\ll\V_2\ll \ldots$ and each varying slowly compared to $\varphi$. To lowest order $\A\V_0=\ri s\varphi'\,\V_0$. Hence $\ri s\varphi'$ and $\V_0$ are respectively the eigenvalues and eigenvectors of $\A$. Equivalently, the $x$-wavenumbers $\kappa_x=-s\varphi'$ are the eigenvalues of $\ri\A$, $\kappa_x=h\,d\varphi/dx=-s\varphi'=\sqrt{s-\kappa ^2}$, $-\sqrt{s-\kappa ^2}$, $s^{1/2} \sec \theta-\kappa\tan\theta\cos\phi$, and $-s^{1/2} \sec \theta-\kappa\tan\theta\cos\phi$, representing respectively the right-moving fast wave, the left-moving fast wave, the rightward Alfv\'en wave, and the leftward Alfv\'en wave. Of course, $\V_0$ is determined by this process only up to a multiplicative scalar function of $s$. To find this scalar we must progress to the next order.

Adapting and completing the method of \citet{wein62} (see his Equation (131)), we have 
\begin{equation}
\left(\A-\ri s\varphi'\I\right)\V_n=s\V_{n-1}'  \,.                                \label{wein_n}
\end{equation}
The amplitude dependence of $\V_0$ on $s$ may then be determined by setting $\V_0=f(s)\boldv_0$, where $\boldv_0$ is an arbitrarily scaled eigenvector of $\A$, and then
premultiplying Equation (\ref{wein_n}), for $n=1$, by $\boldw^T$, where $\boldw$ is the corresponding left eigenvector, thereby extinguishing the left hand side. This leaves $(\ln f)'=-\boldw^T\boldv_0'/\boldw^T\boldv_0$ which may be integrated exactly to find $f(s)$. 

The $\V_n$ $(n\ge1$) may then be calculated recursively by solving Equation (\ref{wein_n}) with constraint $\boldw^T\V_n'=0$. Let $\boldv_n$ be a particular solution of Equation (\ref{wein_n}), with the general solution being $\V_n=\boldv_n+\alpha_n(s)\V_0$. Since $\V_n\ll\V_0$ is required as $s\to\infty$, the constraint fixes $\alpha_n=\int_s^\infty\boldw^T\boldv_n'/\boldw^T\V_0\,\rd s'$, with the integrand assumed a function of the dummy variable $s'$. This integral must be evaluated numerically in practice. Excellent results are attained with the expansion truncated at $n=1$. It is found that $\V_1=\mathcal{O}(s^{-1/2})\V_0$ as $s\to\infty$, so the $\V_1\ll\V_0$ assumption is indeed met on $s\gg1$. 


\subsection{Frobenius about $s=0$}
As $x\to+\infty$ we utilize the Frobenius expansion $\U=\sum_{n=0}^\infty \boldu_n\, s^{n+\mu}$ about $s=0$. This has an infinite radius of convergence since there are no singularities other than $s=0$ itself. Substituting this expansion into Equation (\ref{mateqn}) yields the eigenvalue indicial equation $\A_0\boldu_0=\mu\boldu_0$, thereby specifying both the hitherto unknown index $\mu$ and the zeroth coefficient $\boldu_0$ as the eigenvalues and eigenvectors of $\A_0$ respectively. Later coefficients are found by the recursion $\boldu_n=-(\A_0-(n+\mu)\I)^{-1}\A_1\boldu_{n-1}$. The eigenvalues are $\mu=-\kappa$, $\kappa$, and $\ri\, \kappa\tan\theta\cos\phi$, representing respectively (as $s\to0^+$) the exponentially growing fast wave ($\U_1$), the decaying fast wave ($\U_2$), and an Alfv\'en wave ($\U_3$). Specifically, for the physical fast eigenvalue $\mu=\mu_2=\kappa$, the eigenvector is $\boldu_0^{(2)}=\{-\cos \theta  \cos \phi + \ri \sin \theta ,\sin \phi,-\kappa  \cos \theta  \cos \phi + \ri \kappa  \sin\theta ,\kappa  \sin \phi \}$.

The Alfv\'enic eigenvalue $\mu_3$ has algebraic multiplicity 2 but geometric multiplicity 1.  The only independent eigenvector is $\boldu_0^{(3)} = \{-\ri \cos \theta  \cot \theta  \tan \phi , -\ri \cot\theta ,  \kappa  \cos \theta  \sin \phi , \kappa  \cos\phi \}$. The series expansion has failed to find the full complement of four independent solutions. The fourth solution must therefore take the form
\begin{equation}
\U_4 = \U_3\ln s + \sum_{n=0}^\infty \boldv_n\, s^{n+\mu_3}\, ,
\end{equation} 
where $(\A_0-\mu_3\I)\boldv_0=\boldu_0$ and
\begin{equation}
 \left(\A_0-(n+\mu_3)\I\right)\boldv_n =\left(\boldu_n-\A_1\boldv_{n-1}\right),\quad (n>0)\,.  \label{frobrecur}
\end{equation}
Here the $\boldu_n$ vectors are the coefficients appearing in $\U_3$. The equation for $\boldv_0$ is of course singular and admits an arbitrary additive multiple of $\boldu_0$ in its solution. The later $\boldv_n$ are fully determined by Equation (\ref{frobrecur}) in general.

It remains to determine the combination of $\U_3$ and $\U_4$ (or equivalently the additive multiple of $\U_3$ required in $\U_4$) that corresponds to the pure outgoing Alfv\'en wave. Note that Equation (\ref{basiceqn}) shows that $-\grad_{\!\text{p}}\chi$ is the source term for otherwise free Alfv\'en waves propagating along field lines. As shown by \citet{ca10}, this driving of Alfv\'en waves occurs in a compact mode conversion region near the reflection point, and is characterized by a stationary phase integral. Well to the right of this ($s\ll\kappa$) $\chi$ is both very small and of the wrong phase to interact. To a good approximation therefore, the solution in $s\ll\kappa$ is just that of $\left(\partial_\parallel^2+\omega^2/a^2\right)\bxi  =\mathbf{0}$, \emph{viz.}, $\bxi\propto s^{\ri\kappa\tan\theta\cos\phi}H_0^{(1,2)}(2\sqrt{s}\sec\theta)$, where $H_0^{(1)}$ is the leftward propagating Hankel function (increasing $s$) and $H_0^{(2)}$ moves to the right. We retain the latter only. Now
$H_0^{(2)}(2\sqrt{s}\sec\theta)=1-\ri [2 (\ln \sec \theta +\mathcal{C})+\ln s]/\pi + \mathcal{O}(s) $, where $\mathcal{C}=0.577216\ldots$ is Euler's constant. This is sufficient to determine the radiating Alfv\'en solution, $\U_5=K \U_3+\U_4$, where $K=\ri\pi+\ln\sec^2\theta+2\mathcal{C}-\boldd\vdot\tilde\boldv_0/\boldd\vdot\tilde\boldu_0$. Here, the tilde denotes the restriction of a four-vector to its first two components only, and $\boldd=(\cos\theta\sin\phi,\cos\phi)$ (in $(\perp,y)$ space) is the polarization direction of $\U_4$ as $s\to0$. The idea is to project $\U_3$ and $\U_4$ onto their asymptotic polarization direction, and force the coefficients of 1 and $\ln s$ to appear in the same proportion as in the Hankel function. 
It may be confirmed that $\Div\U_5=0$ to leading order in the Frobenius expansions ($n=0$), but not beyond ($\chi_5=\mathcal{O}(s^{1+\mu_3}\ln s)$), indicating that $\U_5$ is a `pure' Alfv\'en wave only in the limit $s\to0$.

\subsection{Shooting Method}
A bi-directional shooting method is used to solve Equation (\ref{mateqn}) subject to the above boundary conditions. Because of the very different eigenvalues of the four solutions, an automatically switching stiff/nonstiff method is used. Despite this, resolution is lost and the system becomes ill-conditioned as $\kappa$ increases, typically beyond about 8. This limit can be pushed a little higher using high precision arithmetic (20 -- 40 decimal digits), but this soon becomes prohibitively expensive. Thankfully, waves of interest in the solar atmosphere do not usually have wavelengths small compared to the density scale height, so $\kappa\gg1$ is of little practical concern.

\acknowledgements
The authors thank Scott McIntosh for a careful reading of the manuscript and valuable suggestions.


\clearpage


\section*{SUPPLEMENTARY MATERIAL}
Tables 2 to 8 are to be included in `Supplementary Material', not in the paper itself.
\clearpage
\begin{deluxetable}{llllllllllllllll} 
\tabletypesize{\footnotesize}
\rot
\tablewidth{0pt}\tablecolumns{16}
\tablecaption{Forward Alfv\'en conversion coefficient $\mathscr{A}^+$ for a range of $\kappa$ and $\phi$ at magnetic field inclination $\theta=20^\circ$.}
\tablehead{& \multicolumn{15}{c}{Polarization angle $\phi$}\\[4pt]
\colhead{$\kappa$} & \colhead{$5^\circ$} & \colhead{$15^\circ$} & \colhead{$25^\circ$} & \colhead{$35^\circ$} & \colhead{$45^\circ$} & \colhead{$55^\circ$} & \colhead{$65^\circ$} & \colhead{$75^\circ$} & \colhead{$85^\circ$}& \colhead{$95^\circ$} & \colhead{$105^\circ$} & \colhead{$115^\circ$} & \colhead{$125^\circ$} & \colhead{$135^\circ$} & \colhead{$145^\circ$} }
\startdata
0.000 & 0.000 & 0.000& 0.000& 0.000& 0.000& 0.000& 0.000& 0.000& 0.000& 0.000& 0.000& 0.000& 0.000& 0.000& 0.000\\
 0.001 &  0.007 &  0.027 &  0.016 &  0.009 &  0.006 &  0.004 &  0.003 &  0.003 &  0.003 &  0.003 &  0.003 &  0.003 &  0.004 &  0.006 &  0.009 \\
 0.01 &  0.007 &  0.059 &  0.147 &  0.224 &  0.255 &  0.248 &  0.228 &  0.211 &  0.201 &  0.200 &  0.208 &  0.224 &  0.242 &  0.247 &  0.216 \\
 0.1 &  0.008 &  0.067 &  0.177 &  0.321 &  0.476 &  0.619 &  0.730 &  0.797 &  0.813 &  0.782 &  0.708 &  0.602 &  0.476 &  0.344 &  0.220 \\
 0.2 &  0.009 &  0.075 &  0.195 &  0.348 &  0.508 &  0.648 &  0.747 &  0.792 &  0.781 &  0.721 &  0.625 &  0.507 &  0.384 &  0.266 &  0.165 \\
 0.3 &  0.009 &  0.082 &  0.211 &  0.373 &  0.534 &  0.665 &  0.746 &  0.766 &  0.729 &  0.647 &  0.537 &  0.418 &  0.303 &  0.202 &  0.121 \\
 0.4 &  0.010 &  0.088 &  0.226 &  0.394 &  0.554 &  0.676 &  0.737 &  0.733 &  0.672 &  0.573 &  0.457 &  0.340 &  0.237 &  0.152 &  0.088 \\
 0.5 &  0.011 &  0.094 &  0.240 &  0.413 &  0.571 &  0.681 &  0.723 &  0.696 &  0.616 &  0.505 &  0.385 &  0.275 &  0.183 &  0.113 &  0.063 \\
 0.6 &  0.012 &  0.100 &  0.253 &  0.430 &  0.585 &  0.682 &  0.705 &  0.658 &  0.562 &  0.442 &  0.324 &  0.221 &  0.141 &  0.084 &  0.045 \\
 0.7 &  0.012 &  0.106 &  0.265 &  0.445 &  0.596 &  0.681 &  0.685 &  0.620 &  0.510 &  0.386 &  0.271 &  0.177 &  0.108 &  0.062 &  0.032 \\
 0.8 &  0.013 &  0.111 &  0.277 &  0.459 &  0.605 &  0.676 &  0.663 &  0.581 &  0.462 &  0.336 &  0.226 &  0.141 &  0.083 &  0.045 &  0.023 \\
 0.9 &  0.014 &  0.116 &  0.288 &  0.471 &  0.611 &  0.670 &  0.640 &  0.544 &  0.417 &  0.291 &  0.187 &  0.112 &  0.063 &  0.033 &  0.016 \\
 1.0&  0.014 &  0.121 &  0.298 &  0.483 &  0.616 &  0.662 &  0.616 &  0.508 &  0.375 &  0.252 &  0.155 &  0.089 &  0.048 &  0.024 &  0.011 \\
 1.5 &  0.017 &  0.144 &  0.341 &  0.525 &  0.623 &  0.605 &  0.497 &  0.352 &  0.218 &  0.120 &  0.060 &  0.027 &  0.012 &  0.005 &  0.002 \\
 2.0&  0.020 &  0.163 &  0.376 &  0.551 &  0.608 &  0.538 &  0.391 &  0.238 &  0.123 &  0.056 &  0.022 &  0.008 &  0.003 &  0.001 &  0.000 \\
 2.5 &  0.022 &  0.181 &  0.404 &  0.566 &  0.583 &  0.470 &  0.303 &  0.159 &  0.069 &  0.025 &  0.008 &  0.002 &  0.001 &  0.000 &  0.000 \\
 3.0&  0.025 &  0.196 &  0.428 &  0.572 &  0.552 &  0.406 &  0.232 &  0.105 &  0.038 &  0.012 &  0.003 &  0.001 &  0.000 &  0.000 &  0.000 \\
 3.5 &  0.027 &  0.211 &  0.448 &  0.573 &  0.518 &  0.348 &  0.177 &  0.069 &  0.021 &  0.005 &  0.001 &  0.000 &  0.000 &  0.000 &  0.000 \\
 4.0&  0.029 &  0.224 &  0.464 &  0.570 &  0.483 &  0.297 &  0.134 &  0.045 &  0.012 &  0.002 &  0.000 &  0.000 &  0.000 &  0.000 &  0.000 \\
 4.5 &  0.031 &  0.237 &  0.478 &  0.563 &  0.448 &  0.252 &  0.101 &  0.030 &  0.006 &  0.001 &  0.000 &  0.000 &  0.000 &  0.000 &  0.000 \\
 5.0&  0.033 &  0.249 &  0.490 &  0.554 &  0.414 &  0.213 &  0.076 &  0.019 &  0.003 &  0.000 &  0.000 &  0.000 &  0.000 &  0.000 &  0.000 \\

 \enddata 
\label{tab20}
 \end{deluxetable}

\begin{deluxetable}{llllllllllllllll} 
\tabletypesize{\footnotesize}
\rot
\tablewidth{0pt}\tablecolumns{16}
\tablecaption{Forward Alfv\'en conversion coefficient $\mathscr{A}^+$ for a range of $\kappa$ and $\phi$ at magnetic field inclination $\theta=30^\circ$.}
\tablehead{& \multicolumn{15}{c}{Polarization angle $\phi$}\\[4pt]
\colhead{$\kappa$} & \colhead{$5^\circ$} & \colhead{$15^\circ$} & \colhead{$25^\circ$} & \colhead{$35^\circ$} & \colhead{$45^\circ$} & \colhead{$55^\circ$} & \colhead{$65^\circ$} & \colhead{$75^\circ$} & \colhead{$85^\circ$}& \colhead{$95^\circ$} & \colhead{$105^\circ$} & \colhead{$115^\circ$} & \colhead{$125^\circ$} & \colhead{$135^\circ$} & \colhead{$145^\circ$} }
\startdata
0.000 & 0.000 & 0.000& 0.000& 0.000& 0.000& 0.000& 0.000& 0.000& 0.000& 0.000& 0.000& 0.000& 0.000& 0.000& 0.000\\
 0.001 &  0.005 &  0.010 &  0.004 &  0.002 &  0.001 &  0.001 &  0.001 &  0.001 &  0.000 &  0.000 &  0.001 &  0.001 &  0.001 &  0.001 &  0.002 \\
 0.01 &  0.006 &  0.049 &  0.105 &  0.118 &  0.099 &  0.076 &  0.060 &  0.050 &  0.046 &  0.045 &  0.049 &  0.058 &  0.073 &  0.094 &  0.111 \\
 0.1 &  0.007 &  0.060 &  0.158 &  0.283 &  0.411 &  0.520 &  0.591 &  0.619 &  0.611 &  0.573 &  0.513 &  0.435 &  0.343 &  0.246 &  0.156 \\
 0.2 &  0.008 &  0.070 &  0.181 &  0.319 &  0.456 &  0.567 &  0.631 &  0.643 &  0.607 &  0.535 &  0.442 &  0.342 &  0.247 &  0.164 &  0.097 \\
 0.3 &  0.009 &  0.079 &  0.201 &  0.349 &  0.487 &  0.587 &  0.630 &  0.613 &  0.547 &  0.453 &  0.349 &  0.251 &  0.168 &  0.105 &  0.059 \\
 0.4 &  0.010 &  0.087 &  0.220 &  0.375 &  0.511 &  0.597 &  0.615 &  0.571 &  0.482 &  0.374 &  0.269 &  0.181 &  0.113 &  0.066 &  0.035 \\
 0.5 &  0.011 &  0.095 &  0.238 &  0.398 &  0.530 &  0.599 &  0.594 &  0.525 &  0.420 &  0.306 &  0.205 &  0.128 &  0.075 &  0.041 &  0.020 \\
 0.6 &  0.012 &  0.103 &  0.254 &  0.418 &  0.544 &  0.596 &  0.568 &  0.479 &  0.362 &  0.248 &  0.155 &  0.090 &  0.049 &  0.025 &  0.012 \\
 0.7 &  0.013 &  0.110 &  0.269 &  0.436 &  0.554 &  0.589 &  0.540 &  0.434 &  0.310 &  0.199 &  0.117 &  0.063 &  0.032 &  0.015 &  0.007 \\
 0.8 &  0.014 &  0.117 &  0.283 &  0.451 &  0.561 &  0.579 &  0.511 &  0.392 &  0.265 &  0.160 &  0.087 &  0.044 &  0.021 &  0.009 &  0.004 \\
 0.9 &  0.015 &  0.123 &  0.296 &  0.465 &  0.565 &  0.567 &  0.481 &  0.352 &  0.225 &  0.127 &  0.065 &  0.030 &  0.013 &  0.006 &  0.002 \\
 1.0&  0.016 &  0.129 &  0.309 &  0.477 &  0.568 &  0.552 &  0.452 &  0.315 &  0.191 &  0.101 &  0.048 &  0.021 &  0.009 &  0.003 &  0.001 \\
 1.5 &  0.019 &  0.158 &  0.360 &  0.518 &  0.555 &  0.468 &  0.318 &  0.176 &  0.081 &  0.031 &  0.011 &  0.003 &  0.001 &  0.000 &  0.000 \\
 2.0&  0.023 &  0.182 &  0.399 &  0.537 &  0.521 &  0.382 &  0.216 &  0.095 &  0.033 &  0.009 &  0.002 &  0.000 &  0.000 &  0.000 &  0.000 \\
 2.5 &  0.026 &  0.203 &  0.429 &  0.541 &  0.477 &  0.306 &  0.144 &  0.051 &  0.013 &  0.003 &  0.000 &  0.000 &  0.000 &  0.000 &  0.000 \\
 3.0&  0.029 &  0.223 &  0.452 &  0.537 &  0.430 &  0.241 &  0.095 &  0.027 &  0.005 &  0.001 &  0.000 &  0.000 &  0.000 &  0.000 &  0.000 \\
 3.5 &  0.031 &  0.240 &  0.471 &  0.526 &  0.384 &  0.189 &  0.062 &  0.014 &  0.002 &  0.000 &  0.000 &  0.000 &  0.000 &  0.000 &  0.000 \\
 4.0&  0.034 &  0.256 &  0.486 &  0.511 &  0.341 &  0.147 &  0.041 &  0.007 &  0.001 &  0.000 &  0.000 &  0.000 &  0.000 &  0.000 &  0.000 \\
4.5 &  0.037 &  0.271 &  0.497 &  0.494 &  0.301 &  0.114 &  0.026 &  0.004 &  0.000 &  0.000 &  0.000 &  0.000 &  0.000 &  0.000 &  0.000 \\
 5.0&  0.039 &  0.284 &  0.505 &  0.474 &  0.264 &  0.088 &  0.017 &  0.002 &  0.000 &  0.000 &  0.000 &  0.000 &  0.000 &  0.000 &  0.000
\enddata 
\label{tab30}
 \end{deluxetable}

\begin{deluxetable}{llllllllllllllll} 
\tabletypesize{\footnotesize}
\rot
\tablewidth{0pt}\tablecolumns{16}
\tablecaption{Forward Alfv\'en conversion coefficient $\mathscr{A}^+$ for a range of $\kappa$ and $\phi$ at magnetic field inclination $\theta=40^\circ$.}
\tablehead{& \multicolumn{15}{c}{Polarization angle $\phi$}\\[4pt]
\colhead{$\kappa$} & \colhead{$5^\circ$} & \colhead{$15^\circ$} & \colhead{$25^\circ$} & \colhead{$35^\circ$} & \colhead{$45^\circ$} & \colhead{$55^\circ$} & \colhead{$65^\circ$} & \colhead{$75^\circ$} & \colhead{$85^\circ$}& \colhead{$95^\circ$} & \colhead{$105^\circ$} & \colhead{$115^\circ$} & \colhead{$125^\circ$} & \colhead{$135^\circ$} & \colhead{$145^\circ$} }
\startdata
0.000 & 0.000 & 0.000& 0.000& 0.000& 0.000& 0.000& 0.000& 0.000& 0.000& 0.000& 0.000& 0.000& 0.000& 0.000& 0.000\\
 0.001 &  0.004 &  0.005 &  0.002 &  0.001 &  0.000 &  0.000 &  0.000 &  0.000 &  0.000 &  0.000 &  0.000 &  0.000 &  0.000 &  0.000 &  0.001 \\
 0.01 &  0.005 &  0.038 &  0.070 &  0.064 &  0.044 &  0.029 &  0.021 &  0.016 &  0.014 &  0.013 &  0.015 &  0.020 &  0.028 &  0.041 &  0.058 \\
 0.1 &  0.006 &  0.050 &  0.132 &  0.233 &  0.330 &  0.400 &  0.427 &  0.417 &  0.388 &  0.354 &  0.318 &  0.274 &  0.218 &  0.157 &  0.098 \\
 0.2 &  0.007 &  0.061 &  0.158 &  0.275 &  0.386 &  0.465 &  0.496 &  0.479 &  0.426 &  0.355 &  0.278 &  0.204 &  0.139 &  0.087 &  0.049 \\
 0.3 &  0.008 &  0.072 &  0.182 &  0.310 &  0.422 &  0.491 &  0.501 &  0.457 &  0.378 &  0.287 &  0.202 &  0.132 &  0.080 &  0.045 &  0.023 \\
 0.4 &  0.010 &  0.082 &  0.204 &  0.340 &  0.450 &  0.502 &  0.487 &  0.417 &  0.319 &  0.221 &  0.140 &  0.082 &  0.045 &  0.023 &  0.011 \\
 0.5 &  0.011 &  0.091 &  0.224 &  0.367 &  0.470 &  0.504 &  0.464 &  0.373 &  0.264 &  0.167 &  0.095 &  0.050 &  0.025 &  0.011 &  0.005 \\
 0.6 &  0.012 &  0.100 &  0.243 &  0.389 &  0.485 &  0.500 &  0.437 &  0.329 &  0.215 &  0.124 &  0.064 &  0.030 &  0.013 &  0.006 &  0.002 \\
 0.7 &  0.013 &  0.108 &  0.261 &  0.409 &  0.495 &  0.490 &  0.407 &  0.287 &  0.174 &  0.091 &  0.042 &  0.018 &  0.007 &  0.003 &  0.001 \\
 0.8 &  0.014 &  0.116 &  0.277 &  0.426 &  0.501 &  0.478 &  0.377 &  0.249 &  0.139 &  0.067 &  0.028 &  0.011 &  0.004 &  0.001 &  0.000 \\
 0.9 &  0.015 &  0.124 &  0.292 &  0.440 &  0.504 &  0.462 &  0.347 &  0.215 &  0.111 &  0.048 &  0.018 &  0.006 &  0.002 &  0.001 &  0.000 \\
 1.0&  0.016 &  0.131 &  0.306 &  0.453 &  0.505 &  0.445 &  0.318 &  0.184 &  0.088 &  0.035 &  0.012 &  0.004 &  0.001 &  0.000 &  0.000 \\
 1.5 &  0.020 &  0.164 &  0.362 &  0.492 &  0.480 &  0.352 &  0.195 &  0.082 &  0.027 &  0.007 &  0.001 &  0.000 &  0.000 &  0.000 &  0.000 \\
 2.0&  0.024 &  0.192 &  0.404 &  0.504 &  0.433 &  0.265 &  0.115 &  0.035 &  0.008 &  0.001 &  0.000 &  0.000 &  0.000 &  0.000 &  0.000 \\
 2.5 &  0.028 &  0.217 &  0.434 &  0.500 &  0.380 &  0.194 &  0.066 &  0.015 &  0.002 &  0.000 &  0.000 &  0.000 &  0.000 &  0.000 &  0.000 \\
 3.0&  0.032 &  0.238 &  0.457 &  0.487 &  0.328 &  0.140 &  0.038 &  0.006 &  0.001 &  0.000 &  0.000 &  0.000 &  0.000 &  0.000 &  0.000 \\
 3.5 &  0.035 &  0.258 &  0.473 &  0.468 &  0.279 &  0.101 &  0.021 &  0.003 &  0.000 &  0.000 &  0.000 &  0.000 &  0.000 &  0.000 &  0.000 \\
 4.0&  0.038 &  0.275 &  0.485 &  0.446 &  0.236 &  0.071 &  0.012 &  0.001 &  0.000 &  0.000 &  0.000 &  0.000 &  0.000 &  0.000 &  0.000 \\
 4.5 &  0.041 &  0.291 &  0.493 &  0.421 &  0.199 &  0.051 &  0.007 &  0.000 &  0.000 &  0.000 &  0.000 &  0.000 &  0.000 &  0.000 &  0.000 \\
 5.0&  0.044 &  0.306 &  0.498 &  0.396 &  0.166 &  0.036 &  0.004 &  0.000 &  0.000 &  0.000 &  0.000 &  0.000 &  0.000 &  0.000 &  0.000
\enddata 
\label{tab40}
 \end{deluxetable}

\begin{deluxetable}{llllllllllllllll} 
\tabletypesize{\footnotesize}
\rot
\tablewidth{0pt}\tablecolumns{16}
\tablecaption{Forward Alfv\'en conversion coefficient $\mathscr{A}^+$ for a range of $\kappa$ and $\phi$ at magnetic field inclination $\theta=50^\circ$.}
\tablehead{& \multicolumn{15}{c}{Polarization angle $\phi$}\\[4pt]
\colhead{$\kappa$} & \colhead{$5^\circ$} & \colhead{$15^\circ$} & \colhead{$25^\circ$} & \colhead{$35^\circ$} & \colhead{$45^\circ$} & \colhead{$55^\circ$} & \colhead{$65^\circ$} & \colhead{$75^\circ$} & \colhead{$85^\circ$}& \colhead{$95^\circ$} & \colhead{$105^\circ$} & \colhead{$115^\circ$} & \colhead{$125^\circ$} & \colhead{$135^\circ$} & \colhead{$145^\circ$} }
\startdata
0.000 & 0.000 & 0.000& 0.000& 0.000& 0.000& 0.000& 0.000& 0.000& 0.000& 0.000& 0.000& 0.000& 0.000& 0.000& 0.000\\
 0.001 &  0.003 &  0.003 &  0.001 &  0.001 &  0.000 &  0.000 &  0.000 &  0.000 &  0.000 &  0.000 &  0.000 &  0.000 &  0.000 &  0.000 &  0.001 \\
 0.01 &  0.003 &  0.027 &  0.049 &  0.041 &  0.026 &  0.015 &  0.009 &  0.006 &  0.005 &  0.005 &  0.006 &  0.009 &  0.014 &  0.023 &  0.036 \\
 0.1 &  0.005 &  0.039 &  0.102 &  0.179 &  0.249 &  0.289 &  0.285 &  0.250 &  0.211 &  0.185 &  0.169 &  0.151 &  0.122 &  0.086 &  0.052 \\
 0.2 &  0.006 &  0.051 &  0.130 &  0.224 &  0.309 &  0.360 &  0.364 &  0.326 &  0.266 &  0.205 &  0.150 &  0.103 &  0.065 &  0.037 &  0.019 \\
 0.3 &  0.007 &  0.063 &  0.157 &  0.264 &  0.350 &  0.391 &  0.376 &  0.316 &  0.236 &  0.159 &  0.099 &  0.056 &  0.030 &  0.015 &  0.007 \\
 0.4 &  0.009 &  0.074 &  0.182 &  0.298 &  0.382 &  0.406 &  0.367 &  0.285 &  0.192 &  0.114 &  0.060 &  0.029 &  0.013 &  0.006 &  0.002 \\
 0.5 &  0.010 &  0.084 &  0.205 &  0.327 &  0.405 &  0.410 &  0.348 &  0.248 &  0.150 &  0.078 &  0.036 &  0.015 &  0.006 &  0.002 &  0.001 \\
 0.6 &  0.011 &  0.094 &  0.226 &  0.353 &  0.421 &  0.407 &  0.323 &  0.212 &  0.115 &  0.053 &  0.021 &  0.007 &  0.002 &  0.001 &  0.000 \\
 0.7 &  0.012 &  0.104 &  0.246 &  0.374 &  0.432 &  0.398 &  0.296 &  0.179 &  0.087 &  0.035 &  0.012 &  0.004 &  0.001 &  0.000 &  0.000 \\
 0.8 &  0.014 &  0.113 &  0.263 &  0.393 &  0.439 &  0.385 &  0.269 &  0.149 &  0.065 &  0.023 &  0.007 &  0.002 &  0.000 &  0.000 &  0.000 \\
 0.9 &  0.015 &  0.121 &  0.280 &  0.408 &  0.441 &  0.370 &  0.242 &  0.123 &  0.049 &  0.015 &  0.004 &  0.001 &  0.000 &  0.000 &  0.000 \\
 1.0&  0.016 &  0.130 &  0.295 &  0.421 &  0.441 &  0.353 &  0.217 &  0.101 &  0.036 &  0.010 &  0.002 &  0.000 &  0.000 &  0.000 &  0.000 \\
 1.5 &  0.021 &  0.166 &  0.356 &  0.459 &  0.410 &  0.262 &  0.117 &  0.036 &  0.007 &  0.001 &  0.000 &  0.000 &  0.000 &  0.000 &  0.000 \\
 2.0&  0.025 &  0.197 &  0.399 &  0.466 &  0.358 &  0.183 &  0.060 &  0.012 &  0.001 &  0.000 &  0.000 &  0.000 &  0.000 &  0.000 &  0.000 \\
 2.5 &  0.030 &  0.223 &  0.429 &  0.457 &  0.302 &  0.124 &  0.030 &  0.004 &  0.000 &  0.000 &  0.000 &  0.000 &  0.000 &  0.000 &  0.000 \\
 3.0&  0.033 &  0.247 &  0.451 &  0.438 &  0.250 &  0.082 &  0.014 &  0.001 &  0.000 &  0.000 &  0.000 &  0.000 &  0.000 &  0.000 &  0.000 \\
 3.5 &  0.037 &  0.268 &  0.465 &  0.414 &  0.205 &  0.054 &  0.007 &  0.000 &  0.000 &  0.000 &  0.000 &  0.000 &  0.000 &  0.000 &  0.000 \\
 4.0&  0.040 &  0.287 &  0.475 &  0.386 &  0.166 &  0.035 &  0.003 &  0.000 &  0.000 &  0.000 &  0.000 &  0.000 &  0.000 &  0.000 &  0.000 \\
  4.5 &  0.044 &  0.304 &  0.480 &  0.358 &  0.133 &  0.023 &  0.002 &  0.000 &  0.000 &  0.000 &  0.000 &  0.000 &  0.000 &  0.000 &  0.000 \\
 5.0&  0.047 &  0.319 &  0.482 &  0.330 &  0.107 &  0.015 &  0.001 &  0.000 &  0.000 &  0.000 &  0.000 &  0.000 &  0.000 &  0.000 &  0.000
\enddata 
\label{tab50}
 \end{deluxetable}

\begin{deluxetable}{llllllllllllllll} 
\tabletypesize{\footnotesize}
\rot
\tablewidth{0pt}\tablecolumns{16}
\tablecaption{Forward Alfv\'en conversion coefficient $\mathscr{A}^+$ for a range of $\kappa$ and $\phi$ at magnetic field inclination $\theta=60^\circ$.}
\tablehead{& \multicolumn{15}{c}{Polarization angle $\phi$}\\[4pt]
\colhead{$\kappa$} & \colhead{$5^\circ$} & \colhead{$15^\circ$} & \colhead{$25^\circ$} & \colhead{$35^\circ$} & \colhead{$45^\circ$} & \colhead{$55^\circ$} & \colhead{$65^\circ$} & \colhead{$75^\circ$} & \colhead{$85^\circ$}& \colhead{$95^\circ$} & \colhead{$105^\circ$} & \colhead{$115^\circ$} & \colhead{$125^\circ$} & \colhead{$135^\circ$} & \colhead{$145^\circ$} }
\startdata
0.000 & 0.000 & 0.000& 0.000& 0.000& 0.000& 0.000& 0.000& 0.000& 0.000& 0.000& 0.000& 0.000& 0.000& 0.000& 0.000\\
 0.001 &  0.002 &  0.003 &  0.001 &  0.000 &  0.000 &  0.000 &  0.000 &  0.000 &  0.000 &  0.000 &  0.000 &  0.000 &  0.000 &  0.000 &  0.000 \\
 0.01 &  0.002 &  0.017 &  0.035 &  0.033 &  0.020 &  0.010 &  0.005 &  0.003 &  0.002 &  0.002 &  0.003 &  0.005 &  0.009 &  0.017 &  0.027 \\
 0.1 &  0.003 &  0.028 &  0.073 &  0.127 &  0.175 &  0.199 &  0.184 &  0.140 &  0.100 &  0.083 &  0.079 &  0.073 &  0.057 &  0.038 &  0.021 \\
 0.2 &  0.005 &  0.041 &  0.103 &  0.174 &  0.234 &  0.264 &  0.252 &  0.203 &  0.144 &  0.098 &  0.066 &  0.040 &  0.022 &  0.011 &  0.005 \\
 0.3 &  0.006 &  0.053 &  0.132 &  0.217 &  0.281 &  0.300 &  0.268 &  0.201 &  0.128 &  0.072 &  0.037 &  0.017 &  0.007 &  0.003 &  0.001 \\
 0.4 &  0.008 &  0.065 &  0.159 &  0.256 &  0.317 &  0.319 &  0.265 &  0.180 &  0.100 &  0.047 &  0.019 &  0.007 &  0.002 &  0.001 &  0.000 \\
 0.5 &  0.009 &  0.077 &  0.185 &  0.288 &  0.344 &  0.328 &  0.252 &  0.154 &  0.074 &  0.029 &  0.009 &  0.003 &  0.001 &  0.000 &  0.000 \\
 0.6 &  0.011 &  0.088 &  0.208 &  0.316 &  0.363 &  0.327 &  0.233 &  0.128 &  0.053 &  0.017 &  0.004 &  0.001 &  0.000 &  0.000 &  0.000 \\
 0.7 &  0.012 &  0.098 &  0.229 &  0.340 &  0.376 &  0.321 &  0.211 &  0.104 &  0.038 &  0.010 &  0.002 &  0.000 &  0.000 &  0.000 &  0.000 \\
 0.8 &  0.013 &  0.108 &  0.248 &  0.360 &  0.383 &  0.310 &  0.189 &  0.084 &  0.026 &  0.006 &  0.001 &  0.000 &  0.000 &  0.000 &  0.000 \\
 0.9 &  0.014 &  0.117 &  0.266 &  0.376 &  0.386 &  0.296 &  0.167 &  0.066 &  0.018 &  0.003 &  0.000 &  0.000 &  0.000 &  0.000 &  0.000 \\
 1.0&  0.016 &  0.126 &  0.282 &  0.390 &  0.386 &  0.280 &  0.146 &  0.052 &  0.012 &  0.002 &  0.000 &  0.000 &  0.000 &  0.000 &  0.000 \\
 1.5 &  0.021 &  0.165 &  0.347 &  0.427 &  0.354 &  0.197 &  0.070 &  0.015 &  0.002 &  0.000 &  0.000 &  0.000 &  0.000 &  0.000 &  0.000 \\
 2.0&  0.026 &  0.198 &  0.390 &  0.432 &  0.300 &  0.128 &  0.031 &  0.004 &  0.000 &  0.000 &  0.000 &  0.000 &  0.000 &  0.000 &  0.000 \\
 2.5 &  0.030 &  0.227 &  0.421 &  0.419 &  0.245 &  0.081 &  0.013 &  0.001 &  0.000 &  0.000 &  0.000 &  0.000 &  0.000 &  0.000 &  0.000 \\
 3.0&  0.034 &  0.251 &  0.441 &  0.396 &  0.196 &  0.050 &  0.006 &  0.000 &  0.000 &  0.000 &  0.000 &  0.000 &  0.000 &  0.000 &  0.000 \\
 3.5 &  0.038 &  0.273 &  0.454 &  0.369 &  0.154 &  0.030 &  0.002 &  0.000 &  0.000 &  0.000 &  0.000 &  0.000 &  0.000 &  0.000 &  0.000 \\
 4.0&  0.042 &  0.293 &  0.461 &  0.339 &  0.121 &  0.018 &  0.001 &  0.000 &  0.000 &  0.000 &  0.000 &  0.000 &  0.000 &  0.000 &  0.000 \\
  4.5 &  0.046 &  0.311 &  0.465 &  0.310 &  0.093 &  0.011 &  0.000 &  0.000 &  0.000 &  0.000 &  0.000 &  0.000 &  0.000 &  0.000 &  0.000
\enddata 
\label{tab60}
 \end{deluxetable}

\begin{deluxetable}{llllllllllllllll} 
\tabletypesize{\footnotesize}
\rot
\tablewidth{0pt}\tablecolumns{16}
\tablecaption{Forward Alfv\'en conversion coefficient $\mathscr{A}^+$ for a range of $\kappa$ and $\phi$ at magnetic field inclination $\theta=70^\circ$.}
\tablehead{& \multicolumn{15}{c}{Polarization angle $\phi$}\\[4pt]
\colhead{$\kappa$} & \colhead{$5^\circ$} & \colhead{$15^\circ$} & \colhead{$25^\circ$} & \colhead{$35^\circ$} & \colhead{$45^\circ$} & \colhead{$55^\circ$} & \colhead{$65^\circ$} & \colhead{$75^\circ$} & \colhead{$85^\circ$}& \colhead{$95^\circ$} & \colhead{$105^\circ$} & \colhead{$115^\circ$} & \colhead{$125^\circ$} & \colhead{$135^\circ$} & \colhead{$145^\circ$} }
\startdata
0.000 & 0.000 & 0.000& 0.000& 0.000& 0.000& 0.000& 0.000& 0.000& 0.000& 0.000& 0.000& 0.000& 0.000& 0.000& 0.000\\
 0.001 &  0.001 &  0.003 &  0.002 &  0.001 &  0.000 &  0.000 &  0.000 &  0.000 &  0.000 &  0.000 &  0.000 &  0.000 &  0.000 &  0.000 &  0.001 \\
 0.01 &  0.001 &  0.009 &  0.021 &  0.027 &  0.020 &  0.010 &  0.004 &  0.001 &  0.001 &  0.001 &  0.001 &  0.004 &  0.008 &  0.016 &  0.020 \\
 0.1 &  0.002 &  0.018 &  0.047 &  0.081 &  0.111 &  0.127 &  0.116 &  0.077 &  0.041 &  0.030 &  0.031 &  0.027 &  0.018 &  0.010 &  0.005 \\
 0.2 &  0.004 &  0.032 &  0.079 &  0.131 &  0.171 &  0.184 &  0.164 &  0.116 &  0.064 &  0.035 &  0.019 &  0.009 &  0.004 &  0.001 &  0.000 \\
 0.3 &  0.005 &  0.045 &  0.111 &  0.179 &  0.223 &  0.225 &  0.184 &  0.117 &  0.057 &  0.023 &  0.008 &  0.002 &  0.001 &  0.000 &  0.000 \\
 0.4 &  0.007 &  0.058 &  0.141 &  0.221 &  0.264 &  0.251 &  0.188 &  0.106 &  0.043 &  0.013 &  0.003 &  0.001 &  0.000 &  0.000 &  0.000 \\
 0.5 &  0.008 &  0.071 &  0.168 &  0.257 &  0.295 &  0.264 &  0.181 &  0.089 &  0.030 &  0.007 &  0.001 &  0.000 &  0.000 &  0.000 &  0.000 \\
 0.6 &  0.010 &  0.082 &  0.193 &  0.287 &  0.317 &  0.268 &  0.168 &  0.073 &  0.020 &  0.003 &  0.000 &  0.000 &  0.000 &  0.000 &  0.000 \\
 0.7 &  0.011 &  0.093 &  0.215 &  0.312 &  0.332 &  0.264 &  0.152 &  0.058 &  0.013 &  0.002 &  0.000 &  0.000 &  0.000 &  0.000 &  0.000 \\
 0.8 &  0.013 &  0.103 &  0.235 &  0.333 &  0.340 &  0.255 &  0.134 &  0.045 &  0.008 &  0.001 &  0.000 &  0.000 &  0.000 &  0.000 &  0.000 \\
 0.9 &  0.014 &  0.113 &  0.254 &  0.351 &  0.344 &  0.243 &  0.117 &  0.034 &  0.005 &  0.000 &  0.000 &  0.000 &  0.000 &  0.000 &  0.000 \\
 1.0&  0.015 &  0.123 &  0.271 &  0.365 &  0.344 &  0.229 &  0.101 &  0.026 &  0.003 &  0.000 &  0.000 &  0.000 &  0.000 &  0.000 &  0.000 \\
 1.5 &  0.021 &  0.164 &  0.337 &  0.402 &  0.312 &  0.154 &  0.043 &  0.006 &  0.000 &  0.000 &  0.000 &  0.000 &  0.000 &  0.000 &  0.000 \\
 2.0&  0.026 &  0.198 &  0.382 &  0.405 &  0.260 &  0.095 &  0.017 &  0.001 &  0.000 &  0.000 &  0.000 &  0.000 &  0.000 &  0.000 &  0.000 \\
 2.5 &  0.031 &  0.227 &  0.412 &  0.390 &  0.207 &  0.056 &  0.006 &  0.000 &  0.000 &  0.000 &  0.000 &  0.000 &  0.000 &  0.000 &  0.000 \\
 3.0&  0.035 &  0.253 &  0.431 &  0.365 &  0.161 &  0.033 &  0.002 &  0.000 &  0.000 &  0.000 &  0.000 &  0.000 &  0.000 &  0.000 &  0.000 \\
 3.5 &  0.039 &  0.276 &  0.443 &  0.336 &  0.123 &  0.019 &  0.001 &  0.000 &  0.000 &  0.000 &  0.000 &  0.000 &  0.000 &  0.000 &  0.000 \\
 4.0&  0.043 &  0.296 &  0.449 &  0.306 &  0.093 &  0.011 &  0.000 &  0.000 &  0.000 &  0.000 &  0.000 &  0.000 &  0.000 &  0.000 &  0.000
\enddata 
\label{tab70}
 \end{deluxetable}

\begin{deluxetable}{llllllllllllllll} 
\tabletypesize{\footnotesize}
\rot
\tablewidth{0pt}\tablecolumns{16}
\tablecaption{Forward Alfv\'en conversion coefficient $\mathscr{A}^+$ for a range of $\kappa$ and $\phi$ at magnetic field inclination $\theta=80^\circ$.}
\tablehead{& \multicolumn{15}{c}{Polarization angle $\phi$}\\[4pt]
\colhead{$\kappa$} & \colhead{$5^\circ$} & \colhead{$15^\circ$} & \colhead{$25^\circ$} & \colhead{$35^\circ$} & \colhead{$45^\circ$} & \colhead{$55^\circ$} & \colhead{$65^\circ$} & \colhead{$75^\circ$} & \colhead{$85^\circ$}& \colhead{$95^\circ$} & \colhead{$105^\circ$} & \colhead{$115^\circ$} & \colhead{$125^\circ$} & \colhead{$135^\circ$} & \colhead{$145^\circ$} }
\startdata
0.000 & 0.000 & 0.000& 0.000& 0.000& 0.000& 0.000& 0.000& 0.000& 0.000& 0.000& 0.000& 0.000& 0.000& 0.000& 0.000\\
 0.001 &  0.000 &  0.002 &  0.003 &  0.002 &  0.001 &  0.000 &  0.000 &  0.000 &  0.000 &  0.000 &  0.000 &  0.000 &  0.000 &  0.001 &  0.002 \\
 0.01 &  0.000 &  0.003 &  0.007 &  0.012 &  0.015 &  0.013 &  0.006 &  0.001 &  0.000 &  0.000 &  0.001 &  0.005 &  0.009 &  0.009 &  0.007 \\
 0.1 &  0.001 &  0.012 &  0.029 &  0.049 &  0.063 &  0.068 &  0.061 &  0.041 &  0.013 &  0.007 &  0.006 &  0.003 &  0.001 &  0.000 &  0.000 \\
 0.2 &  0.003 &  0.026 &  0.064 &  0.103 &  0.129 &  0.129 &  0.102 &  0.059 &  0.020 &  0.006 &  0.001 &  0.000 &  0.000 &  0.000 &  0.000 \\
 0.3 &  0.005 &  0.040 &  0.098 &  0.155 &  0.188 &  0.178 &  0.129 &  0.064 &  0.018 &  0.003 &  0.000 &  0.000 &  0.000 &  0.000 &  0.000 \\
 0.4 &  0.006 &  0.054 &  0.129 &  0.200 &  0.233 &  0.210 &  0.140 &  0.061 &  0.013 &  0.001 &  0.000 &  0.000 &  0.000 &  0.000 &  0.000 \\
 0.5 &  0.008 &  0.067 &  0.158 &  0.238 &  0.267 &  0.227 &  0.139 &  0.052 &  0.008 &  0.000 &  0.000 &  0.000 &  0.000 &  0.000 &  0.000 \\
 0.6 &  0.010 &  0.079 &  0.183 &  0.269 &  0.290 &  0.233 &  0.131 &  0.042 &  0.005 &  0.000 &  0.000 &  0.000 &  0.000 &  0.000 &  0.000 \\
 0.7 &  0.011 &  0.090 &  0.206 &  0.295 &  0.306 &  0.231 &  0.118 &  0.033 &  0.003 &  0.000 &  0.000 &  0.000 &  0.000 &  0.000 &  0.000 \\
 0.8 &  0.012 &  0.101 &  0.227 &  0.317 &  0.315 &  0.223 &  0.104 &  0.025 &  0.002 &  0.000 &  0.000 &  0.000 &  0.000 &  0.000 &  0.000 \\
 0.9 &  0.014 &  0.111 &  0.246 &  0.335 &  0.319 &  0.212 &  0.090 &  0.018 &  0.001 &  0.000 &  0.000 &  0.000 &  0.000 &  0.000 &  0.000 \\
 1.0&  0.015 &  0.120 &  0.263 &  0.349 &  0.319 &  0.200 &  0.076 &  0.013 &  0.000 &  0.000 &  0.000 &  0.000 &  0.000 &  0.000 &  0.000 \\
 1.5 &  0.021 &  0.162 &  0.331 &  0.386 &  0.288 &  0.130 &  0.030 &  0.002 &  0.000 &  0.000 &  0.000 &  0.000 &  0.000 &  0.000 &  0.000 \\
 2.0&  0.026 &  0.198 &  0.376 &  0.388 &  0.236 &  0.078 &  0.011 &  0.000 &  0.000 &  0.000 &  0.000 &  0.000 &  0.000 &  0.000 &  0.000 \\
 2.5 &  0.031 &  0.228 &  0.406 &  0.372 &  0.185 &  0.044 &  0.004 &  0.000 &  0.000 &  0.000 &  0.000 &  0.000 &  0.000 &  0.000 &  0.000 \\
 3.0&  0.035 &  0.254 &  0.425 &  0.346 &  0.141 &  0.024 &  0.001 &  0.000 &  0.000 &  0.000 &  0.000 &  0.000 &  0.000 &  0.000 &  0.000 \\
 3.5 &  0.040 &  0.277 &  0.436 &  0.317 &  0.106 &  0.013 &  0.000 &  0.000 &  0.000 &  0.000 &  0.000 &  0.000 &  0.000 &  0.000 &  0.000
 \enddata 
\label{tab80}
 \end{deluxetable}

\end{document}